\documentclass[aps,prl,twocolumn,superscriptaddress,a4paper,nopacs]{revtex4-1}
\pdfoutput=1

\usepackage{graphicx}
\usepackage{amsmath}

\usepackage[colorlinks=true,citecolor=blue,linkcolor=magenta]{hyperref}
\hypersetup{
pdfauthor = {P. Maletinsky},
urlcolor=blue, bookmarksnumbered =  true}

\usepackage{graphicx}

\begin{document}

\title{Color centers in diamond as novel probes of superconductivity}

\author{Victor M.\ Acosta}
\affiliation{Center for High Technology Materials, University of New Mexico, 1313 Goddard St.\ SE, Albuquerque, NM 87106, USA}
\author{Louis S.\ Bouchard}
\affiliation{Department of Chemistry and Biochemistry, University of California Los Angeles, 607 Charles E. Young Drive East, Los Angeles, CA 90095, USA}
\author{Dmitry Budker}
\email{budker@uni-mainz.de}
\affiliation{Helmholtz Institut, Johannes Gutenberg-Universit\"at Mainz, 55099 Mainz, Germany}
\affiliation{Department of Physics, University of California, Berkeley, California 94720-7300, USA}
\author{Ron Folman}
\affiliation{ Department of Physics, Ben-Gurion University of the Negev, Be'er Sheva 84105, Israel}
\author{Till Lenz}
\affiliation{Johannes Gutenberg-Universit\"at Mainz, 55128 Mainz, Germany}
\author{Patrick Maletinsky}
\affiliation{Department of Physics, University of Basel, Klingelbergstrasse 82, Basel CH-4056, Switzerland}
\author{Dominik Rohner} 
\affiliation{Department of Physics, University of Basel, Klingelbergstrasse 82, Basel CH-4056, Switzerland}
\author{Yechezkel Schlussel}
\affiliation{Physikalisches Institut and Center for Quantum Science (CQ) in LISA$^+$, Universit\"at T\"ubingen, Auf der Morgenstelle 14, D-72076 T\"ubingen , Germany}
\author{Lucas Thiel}
\affiliation{Department of Physics, University of Basel, Klingelbergstrasse 82, Basel CH-4056, Switzerland}

\date{\today}

\begin{abstract}
Magnetic imaging using color centers in diamond through both scanning and wide-field methods offers a combination of unique capabilities for studying superconductivity, for example, enabling accurate vector magnetometry at high temperature or high pressure, with spatial resolution down to the nanometer scale. The paper briefly reviews various experimental modalities in this rapidly developing nascent field and provides an outlook towards possible future directions.
\end{abstract}

\maketitle

\section{Introduction}
\label{sec:Intro}

Many experimental techniques exist for probing superconductors (SC) \cite{Bending1999}; however, the relatively new technology based on color centers in diamond offers a set of unique capabilities to help solve some of the open questions in the physics of high-critical-temperature (T$_{\rm{c}}$)  SC.  These capabilities include possible combinations of features such as operation across a wide temperature range (from millikelvin to above room temperature), the ability to obtain spatial resolution down to the nanometer scale, vector-magnetometry capabilities, and the possibility to perform dynamic (AC) sensing up to the GHz range. The open questions that may be addressed with diamond-based methods include measuring the detailed structure of flux vortices, investigation of vortex dynamics, and the study of percolation effects and phenomena such as ``island superconductivity'' \cite{Kresin2006,Wolf2012} or superconductivity in metal clusters \cite{Kresin2008,Kresin2012,Kresin2013}.

In this review, we provide a brief introduction to the physics of the nitrogen-vacancy color centers in diamond and various magnetometry modalities based on this system. We then review the work that has been done up until now on applications to the study of superconductors and discuss possible future directions.

We note that another context where color centers in diamond and superconductivity are frequently discussed together is the development of hybrid quantum circuits for information processing (see, for example, the review \cite{AMSUSS2014}). However, it is not the intent of this article to discuss such hybrid quantum  systems.

For the convenience of the reader, Table\,\ref{Table:abbreviations} lists some of the acronyms used in this paper and their meaning.
\begin{figure}[t!]
\centering
\includegraphics{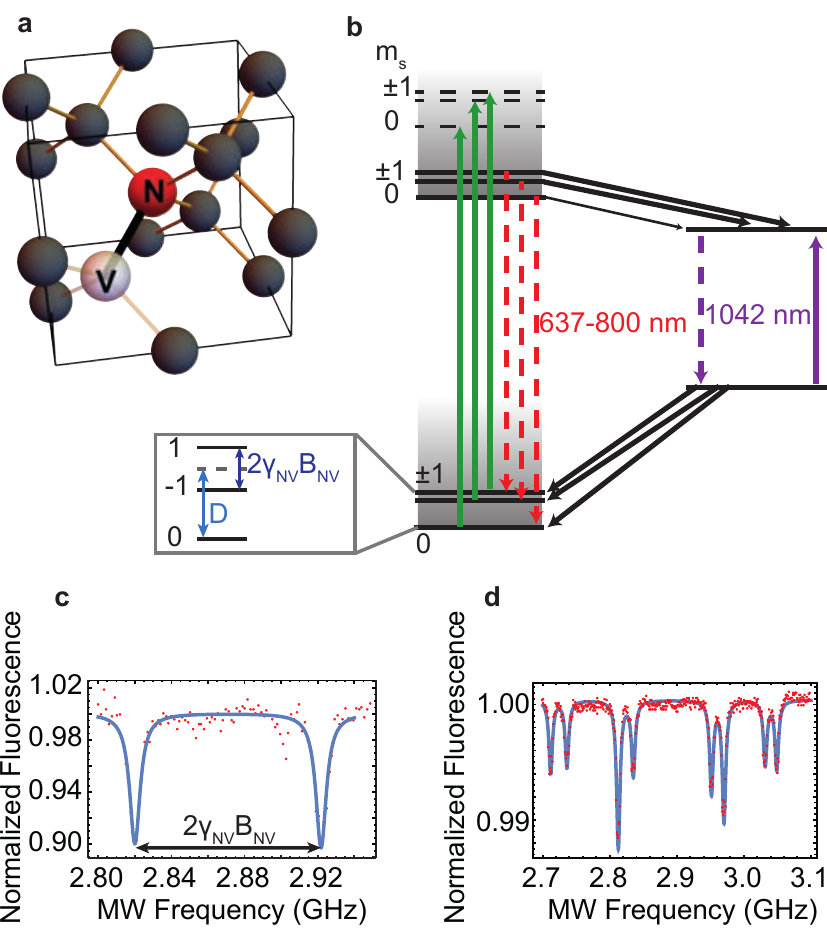}
\caption{a) Diamond unit cell with an incorporated nitrogen atom (red) next to a vacancy (white), forming an NV center. b) Energy-level diagram of the NV center. Two triplet states with an energy separation corresponding to the energy of a photon with $\approx637$\,nm wavelength and two singlet states located between the triplet states, whose separation corresponds to $\approx$1042\,nm. In the triplet ground state, the $m_{s}=\pm1$ sublevels are degenerate and separated by $D$ from the $m_{s}=0$ sublevel in the absence of symmetry-breaking fields and strain. A splitting of the $m_{s}=1$ and $m_{s}=-1$ state is induced by the magnetic fields, allowing for magnetometry using electron spin resonance. c) ODMR spectrum of a single NV center, which allows for the determination of the B-field along the NV axis. d) ODMR spectrum of an ensemble of NV centers, with a field applied along an arbitrary direction, such that four different B-field projections on the different NV axes are observable. Analyzing such spectra, it is possible to reconstruct the full B-field vector.}
\label{levelandstructure}
\end{figure}

\begin{table}[t]
\caption{Acronyms and their meanings.}
\medskip 
\center
\begin{tabular}{ll}
\hline \hline
Acronym~~ & Meaning \\
\hline

\rule{0ex}{2.6ex} AFM & atomic-force microscope \\
\rule{0ex}{2.6ex} BSCCO & bismuth-strontium-calcium-copper-oxide \\
\rule{0ex}{2.6ex} CFB & coherent (image-preserving) fiber bundle \\
\rule{0ex}{2.6ex} NV & nitrogen-vacancy (center in diamond) \\
\rule{0ex}{2.6ex} ODMR & optically detected magnetic resonance \\
\rule{0ex}{2.6ex} PSB & phonon sideband \\
\rule{0ex}{2.6ex} SC & superconductor, superconducting \\
\rule{0ex}{2.6ex} STED & stimulated-emission depletion (microscopy) \\
\rule{0ex}{2.6ex} YBCO & yttrium barium copper oxide \\
\rule{0ex}{2.6ex} ZPL & zero-phonon line \\

\hline \hline
\end{tabular}
\label{Table:abbreviations}
\end{table}

\section{NV centers in diamond}
\label{sec:NVcenters}

The negatively charged nitrogen-vacancy (NV) center, further referred to as the NV center, is one of the most studied color centers in diamond.
The NV center itself comprises a substitutional nitrogen atom and a neighboring vacancy, as well as an additionally captured electron. The line going through the N and the V (NV axis) is the symmetry axis of the system and can be aligned along four different crystallographic axes (see Fig.\ref{levelandstructure}a).
Recently, the NV center gained interest due to its remarkable properties as a nanoscale sensor \cite{Chernobrod2005,Taylor2008,Degen2008}. It was shown that it can be used to sense, for example, temperature, magnetic and electric fields \cite{TempBudker,Acosta2010a,Dolde2011}. Here, we focus on the application as magnetic-field sensor since this is the quantity of primary interest when investigating superconductors with NV centers.
For sensitive magnetometry with NV centers one makes use of their unique energy-level scheme (see Fig.\ref{levelandstructure}b). The NV center in its ground electronic state is a spin-1 triplet. The excited triplet state is separated from the ground state by an electronic energy corresponding to light of $\approx\,637$\,nm wavelength and can be excited into the phonon sideband using green light (usually 532\,nm or 515\,nm). After excitation to the upper triplet state, the center mainly relaxes towards the ground electronic state and its vibronic sideband by emitting a photon in the range of $\approx\,$637-800\,nm.

The $m_{s}=\pm1$ magnetic sublevels in the ground state are separated from the $m_{s}=0$ sublevel by $D$, where 
$D \approx 2.87$\,GHz.
For a magnetic field $\mathbf{B}_{\rm{NV}}$ applied along the NV axis, the $m_{s}=\pm1$ sublevels split due to the Zeeman effect by $\Delta\nu=2\gamma_{e}\mathbf{B}_{\rm{NV}}\cdot\mathbf{S}$. Here, $\gamma_{e}$ is the electron gyromagnetic ratio and $\mathbf{S}$ is the spin of the NV center. It is important to note that the presence of crystal strain and electric fields may have a significant effect on the electronic structure of the NV center, see, for example, a detailed discussion in \cite{Rondin2014}.

The amount of fluorescence emitted by an NV center is dependent on its spin state. For a single NV center, the fluorescence rate for the $m_{s}=\pm1$ states is $\approx\,30\,\%$ lower than the fluorescence rate of the $m_{s}=0$ state \cite{Jelezko2006a}. This is due to spin-dependent non-radiative decay towards the upper  of the two singlet states (see Fig.\ref{levelandstructure}b), from which the NV center decays (mostly nonradiatively) into the lower singlet state, which has an $\approx\,40$ times longer lifetime than the upper triplet state \cite{Robledo2011NJP,Dumeige2013}. From the lower singlet state, there is significant probability for a decay into each of the $m_{s}$ sublevels (see \cite{Kalb2018}). The much higher probability for the $m_s=\pm1$ states to ``leak'' into the singlets also leads to the fact that the spin can be initialized to the $m_{s}=0$ state when pumped with green light \cite{Jelezko2006a}. Due to these properties, the magnetic field can be determined by optically-detected magnetic resonance (ODMR) spectroscopy. To obtain an ODMR spectrum, a microwave field is applied to the NV center under, for instance, continuous green illumination and fluorescence is detected. When the microwave frequency matches the transition frequency between the $m_{s}=0$ and one or both of the $m_{s}=\pm1$ states, the population is transferred to the corresponding state(s) resulting in a reduced fluorescence signal. By determining the transition frequency one can then determine the magnetic field (see Fig.\,\ref{levelandstructure}d).
For magnetometry as described below, it is important that for measurements conducted in the presence of a bias magnetic field in excess of a few mT, this field should be aligned with the axis of the NV centers used because transverse magnetic fields lead to quenching of the NV fluorescence and loss of ODMR contrast\,\cite{Tetienne2012}.  Further details on the physics of the NV centers can be found, for example, in Refs.\,\cite{Jelezko2006a,Rondin2014,Doherty2013,Jensen2017}.
\begin{figure}[t!]
\centering
\includegraphics{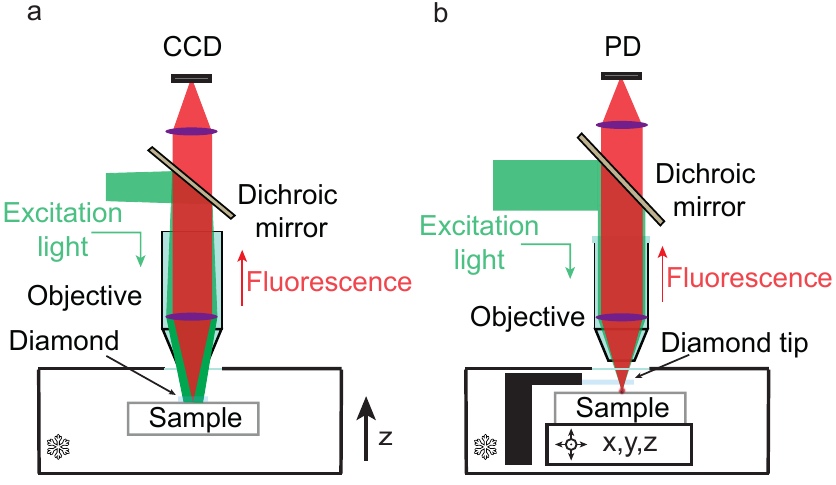}
\caption{a) Schematic of a wide-field-imaging NV setup. An area of the diamond is illuminated with green light; red fluorescent light is transmitted through the dichroic mirror and is projected onto a camera (for instance, a charge-coupled device, CCD). b) Scanning-probe single-NV setup. The green laser is focused on a single NV center located in an all-diamond AFM tip and the fluorescence is detected with a single-pixel photodetector (PD). To acquire an image, the position of the sample below the tip is scanned.}
\label{setups}
\end{figure}

In this paper, two different approaches of NV center based magnetometry are described, which have both been used to study vortices in type-II superconductors. The first one is NV-ensemble magnetometry, where many NV centers are observed at the same time and the second is scanning-probe single-NV magnetometry, where a single NV center is used to measure the local magnetic field at the NV center's position \cite{Taylor2008}.

In the first approach based on ensembles of NV centers, one usually uses a diamond plate with a high density layer of NV centers close to the diamond surface. The diamond is placed on the SC and a wide-field microscope is used to image the magnetic field (see Fig.\,\ref{setups}a). This has the advantage that a large area can be imaged at the same time without scanning the beam, sample or probe. In addition, the acquisition time is typically reduced due to the higher fluorescence emitted by multiple NV centers. On the other hand, the resolution in this approach is usually limited by the optical diffraction limit which is on the order of $400\,$nm. (We note that better resolution, $\approx\,150\,$nm FWHM, may be obtained with solid-immersion lenses \cite{Siyushev2010} or using wide-field super-resolution methods. Resolution better then $10\,$nm is theoretically possible  \cite{Gustafsson2005,Chmyrov2013}, though super resolution  has not been demonstrated with NV magnetic microscopy yet.)

Due to the atomic size of the NV, scanning-probe single-NV magnetometry (see Fig.\,\ref{setups}b) can be used for magnetometry with a resolution of nanometers \cite{Thiel2016,Pelliccione2016} (and down to below 1\,nm with the use of magnetic gradients \cite{Grinolds2014}).  A single NV center is positioned in the focus of a confocal microscope and the sample is scanned in position below the NV center to acquire an image. To bring the NV center as close as possible to the sample, all-diamond atomic-force-microscope (AFM) tips were developed, which contain a single NV center within the first few nm of the tip \cite{fabricationMaletinsky}. In most cases, the achieved spatial resolution is limited by the distance between the imaged surface and the NV center.

Both approaches are highly relevant for the study of SC because of their different advantages and are described in more detail in the following sections.

\section{Early work: probing bulk properties of superconductors}
\label{sec:EarlyWork}

The earliest studies using diamond magnetometry to  probe superconductors involved detection of the Meissner effect in bulk superconductors. In the superconducting state, materials exhibit perfect diamagnetism. Upon application of an external magnetic field, flux is expelled from the interior of the superconductor, resulting in perturbations in the stray magnetic fields around the superconductor. 

\begin{figure}[t]
    \includegraphics[width=\columnwidth]{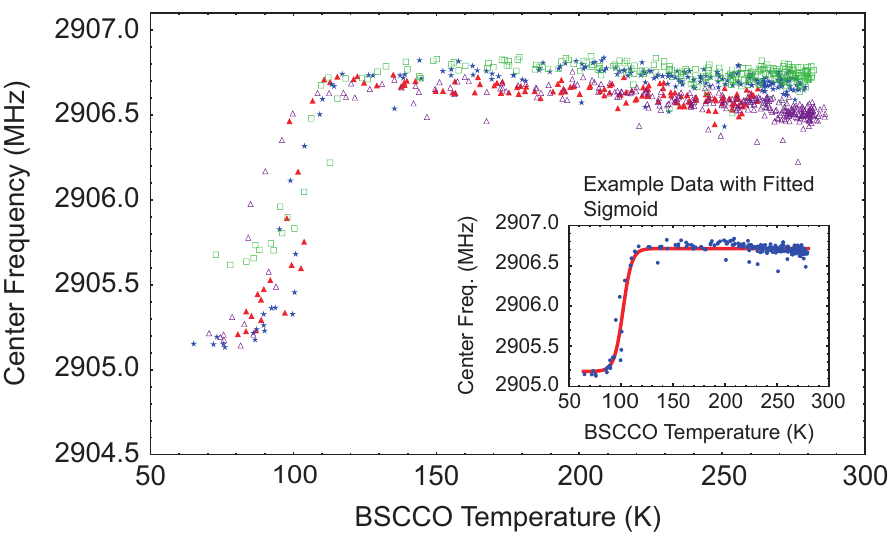}
    \caption{A BSCCO-2223 sample showed a phase transition to the superconducting state at T$_{\rm{c}}$=102(3) K as measured via monitoring the change of the resonance frequency of an NV-center ensemble. The four symbol types represent repeated runs on the same sample performed at different times. (Adapted from \cite{bouchard2011}.)}
    \label{fig:Bouchard2011_Fig4}
    \end{figure}

The first study using diamond magnetometry to probe the Meissner effect \cite{bouchard2011} involved placing a bismuth strontium calcium copper oxide (BSCCO) crystal in close proximity to a diamond plate uniformly doped  with NV centers. The BSCCO crystals were submerged in liquid nitrogen, then placed on top of the diamond magnetometer and allowed to warm to room temperature. Meanwhile, the NV ODMR spectrum was recorded continuously, and the resulting central frequencies measured the change in the stray magnetic field produced by the BSCCO crystal. Figure\,\ref{fig:Bouchard2011_Fig4} shows results of these measurements. Clear signatures of the superconducting transition are observed with a measured $T_c=102\pm 6~{\rm K}$, in good agreement with the value of 105\,K measured using other techniques \cite{Lim1997}.

While performing these early experiments, a significant temperature dependence of the NV zero-field splitting parameter, $D$, was observed. When monitoring a single ODMR frequency (e.g., that of the $m_s=0\leftrightarrow m_s=-1$ transition) in variable-temperature studies, the effect of $dD/dT$ is indistinguishable from changes in the magnetic field. This effect was subsequently investigated in detail experimentally \cite{Acosta2010,Chen2011} and theoretically \cite{Doherty2014}, and it was found that $dD/dT$ ranges from $\approx\,-74$\,kHz/K at room temperature to $\approx\,-35\,$kHz/K at 200\,K, to $-dD/dT<$6\,kHz/K at 100\,K and below, where  such temperature shifts become, in most cases, irrelevant. The temperature dependence of $D$ was used to image thermal effects in condensed matter \cite{Laraoui2015} and living systems \cite{Kucsko2013}. 
It could also be employed to perform thermometry, together with magnetometry, for superconductors at temperatures $>$100\,K. In order to unambiguously disentangle changes in temperature from changes in magnetic field and thus simultaneously measure these quantities, several protocols have been pursued, including monitoring both $m_s=0\leftrightarrow m_s=\pm1$ transitions simultaneously \cite{Kehayias2014} or directly monitoring the $m_s=-1\leftrightarrow m_s=+1$ transition using double-quantum pulse protocols \cite{Fang2013}. 

\section{Wide-field imaging of superconducting vortices}
\label{sec:WideField}

\begin{figure}[t]
    \includegraphics[width=\columnwidth]{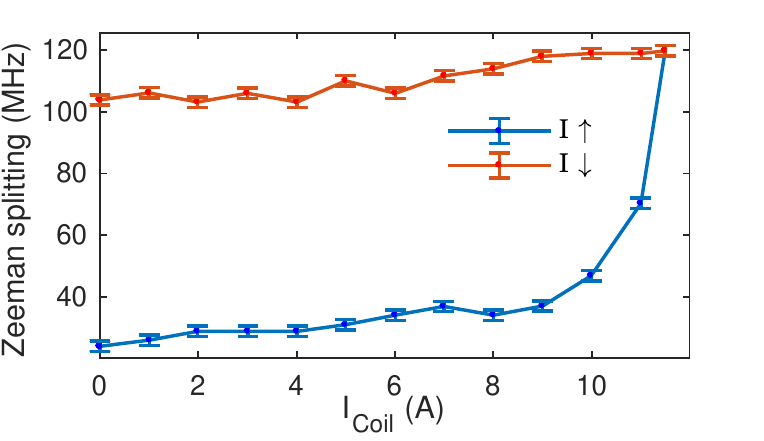}
    \caption{
The Zeeman splitting of the energy levels (from which one can retrieve the magnetic field) of NV centers positioned over an SC (YBCO) layer. Measurements were carried out while increasing and decreasing the external magnetic field (proportional to the coil current, I). The plot demonstrates the local transition from the Meissner state to an intermediate state in which vortices appear. Blue points correspond to increasing coil current while red points correspond to decreasing current (from \cite{PRBwaxman}).}
    \label{fig:Waxman2014PenetrationField}
    \end{figure}

Several magnetic studies of superconducting phenomena have already been carried out using 
    ensemble NV-based magnetometry.
    Waxman et al.  measured the SC-normal phase transition and the average field of vortices in an YBCO thin-film layer \cite{PRBwaxman} (Fig.~\ref{fig:Waxman2014PenetrationField}).

In this experiment, a $1\,\mathrm{mm}\times2\,\mathrm{mm}\times80\,\mu\mathrm{m}$ diamond with a thin ($15\,\mathrm{nm}$) dense (a few $\times 10^{11}$\, cm$^{-2}$ surface density) layer of NV centers positioned on the superconductor was used. Although this method enabled the study of several magnetic phenomena, it turned out to be insufficient in terms of the spatial resolution required for vortex imaging. This spatial resolution was limited mainly by the NV-superconductor distance.

Another limitation of the setup described in \cite{PRBwaxman} came from the use of large windows of the optical cryostat which led to excess heating and made it difficult to reach temperatures below $10\,\mathrm{K}$. To overcome such limitations, Alfasi et al. glued a diamond to a coherent (i.e., image preserving) fiber bundle (CFB) with 30,000 cores and used it in a liquid-helium environment\,\cite{Alfasi2016}. In this experiment, the fluorescence signal from the NV centers was collected with the coherent fiber bundle and projected onto a CCD. Alfasi et al. studied the shielding currents in a Nb stripe in the presence of different external magnetic fields by measuring changes in the ODMR spectra across the edge of the SC and comparing them to theoretical predictions. From this comparison, they concluded that the distance of the sensor from the superconductors was approximately $7\,\mu\mathrm{m}$, which was attributed to techniques used to generate the NV centers\,\cite{Alfasi2016}. Apart from this, when using CFB, spatial resolution is limited by the diameter of a single fiber core and by the distance between the different fibers, which is typically $5\,\mu \mathrm{m}$.

\begin{figure}[t]
    \includegraphics[width=\columnwidth]{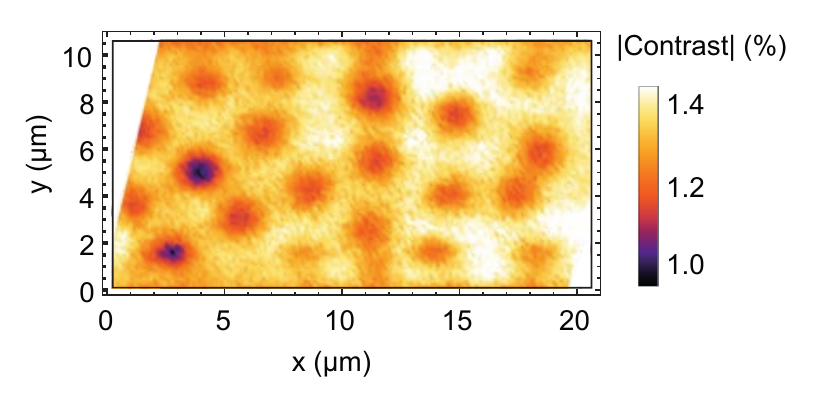}
    \caption{Pinned vortices in a YBCO thin film after cooling it at 0.18\,mT. The image was generated using wide-field NV center based magnetometry (adapted from\,\cite{schlussel2018widefield}).}
    \label{fig:Till2018vortices}
    \end{figure}

To achieve both a small sample-sensor distance, and a high acquisition rate for the imaging of vortices, Schlussel et al. used $20\,\mu \mathrm{m}\times10\,\mu \mathrm{m}\times2\,\mu \mathrm{m}$ single-crystal diamond plates positioned on a thin-film YBCO chip. The guiding idea was that such thin diamond membranes are deformable and allow for adhesion to the SC substrate even in the presence of surface imperfections.
In this work, different densities of vortices as a function of the cooling field were imaged. An example of a magnetic map that images pinned vortices is shown in Fig.~\ref{fig:Till2018vortices}. By comparing the magnetic field measured by the NV sensor to the predicted magnetic field above a vortex, the distance between the superconductor and the magnetic sensor was estimated to be $\approx\,550\,\mathrm{nm}$. 
    This distance still limits the spatial resolution and reduces the acquisition rate.

Recently, Nusran et al. used a similar setup to the one in Fig.\ref{setups}b, but in a confocal microscope instead of a wide-field microscope. They attached mm sized diamond plates with a near-surface NV layer to different SC\,\cite{Nusran2018}. They scanned both SC and diamond together and observed the Meissner effect in different SCs with a resolution of $\approx\,500$\,nm in the x-y direction given by resolution of the used confocal microscope and $\approx\,20$\,nm in z direction given by the thickness of the NV layer. Using this technique they were able to observe conventional Meissner effect in single-crystal Pb as well as an unusual paramagnetic Meissner effect together with random variation of the local magnetic field in a single-crystal Nb sample. Later on, Joshi et al. used the same system to better measure the lower critical fields of several type-II superconductors \cite{Joshi2018}.

Using a wide-field configuration to image magnetic fields with NV centers carries a disadvantage of higher required laser power (depending on the size of the illuminated area) to excite the NV centers.
Since superconductivity requires cryogenic temperatures, heat transfer to the SC is a concern.
Local heating around the position of the laser spot mainly has two effects. Firstly, it can heat the sample above the critical temperature and therefore destroy the superconductivity locally. This will be especially harmful if low-T$_{\rm{c}}$ SC are studied since it might be difficult to achieve such low temperatures. Secondly, it makes it difficult to determine the exact temperature of the sample at the investigated spot. This problem could be solved (for high-T$_{\rm{c}}$ SC) by using the temperature dependence of $D$ \cite{TempBudker}. Heating of the SC can be reduced by performing pulsed measurements \cite{Dreau2011} at low duty cycles, such that the transferred energy/heat per unit time is reduced. Heating can also be potentially reduced by using resonant excitation on the zero-phonon line (ZPL), which has larger optical-absorption cross section than the phonon sideband (PSB) at low temperature. A possible complication is more efficient NV photoionization compared with the green light. Finally, there is a possibility to use either a reflectively coated diamond plate or SC, which would reduce the amount of absorbed light. In this case one has to pay the price of increased minimum NV-SC distance ($\approx\,100\,$nm in case of silver coatings). 

In the past decade, apart from conventional ODMR, many different protocols for both AC and DC magnetic sensing using NV centers have been proposed and demonstrated \cite{Casola2018,Jensen2017,Rondin2014}. These different protocols have allowed optimization of the magnetic sensitivity and spatial resolution of NV-based magnetometers. In addition, control over the NV-center concentrations \cite{Lesik2016}, which is important for the sensitivity of NV-based magnetometers, has improved. Also, diamond crystals with preferential orientation of NV-center ensembles have been grown \cite{Lesik2014,Lesik2015}. In this case, only one of the four possible NV alignments (see Fig.\,\ref{levelandstructure}a) is populated, which improves magnetic sensitivity since only fluorescence from the one selected NV alignment relevant for magnetometry is collected. For reference, typical sensitivity of present systems is $\approx\,1\,\mu$T\,s$^{1/2}$ per 300\,nm$^2$ diffraction-limited spot \cite{Kleinsasser2016}.

\section{Nanoscale quantum sensing of superconductors}
\label{sec:ScanningProbe}

Superconductors exhibit magnetic inhomogeneities down to the nanoscale, either due to impurities, changes in local structure, or vortex effects. 
Addressing nanoscale properties of superconducting samples requires approaches which go beyond the wide-field imaging experiments described in Sec.\,\ref{sec:WideField}. In principle, as discussed in Sec.\,\ref{sec:NVcenters}, wide-field imaging can be extended to sub-$100~$nm resolution by combining recently developed super-resolution methods such as stimulated-emission depletion (STED)\,\cite{Rittweger2009} with wide-field imaging. While potentially a viable approach, such wide-field super-resolution imaging has thus far not been demonstrated in cryogenic environments, in part due to the prohibitively high laser intensities of up to MW/cm$^2$ (depending on the specifics of the method used) required to achieve nanoscale resolution\,\cite{Rittweger2009}. Alternatively, single NV centers can be employed to perform nanoscale magnetic imaging in a scanning probe setting\,\cite{Rondin2014}, where the NV is scanned within several tens of nanometers distance from a sample to yield nanoscale magnetic-field maps. This approach has been successfully applied in the past to image ferro-\,\cite{Tetienne2014a} and antiferromagnetic\,\cite{Gross2017} spin-textures and has recently been adopted to cryogenic conditions to yield quantitative, nanoscale images of vortex stray fields\,\cite{Thiel2016}.

\begin{figure}[t]
  \includegraphics[width=\columnwidth]{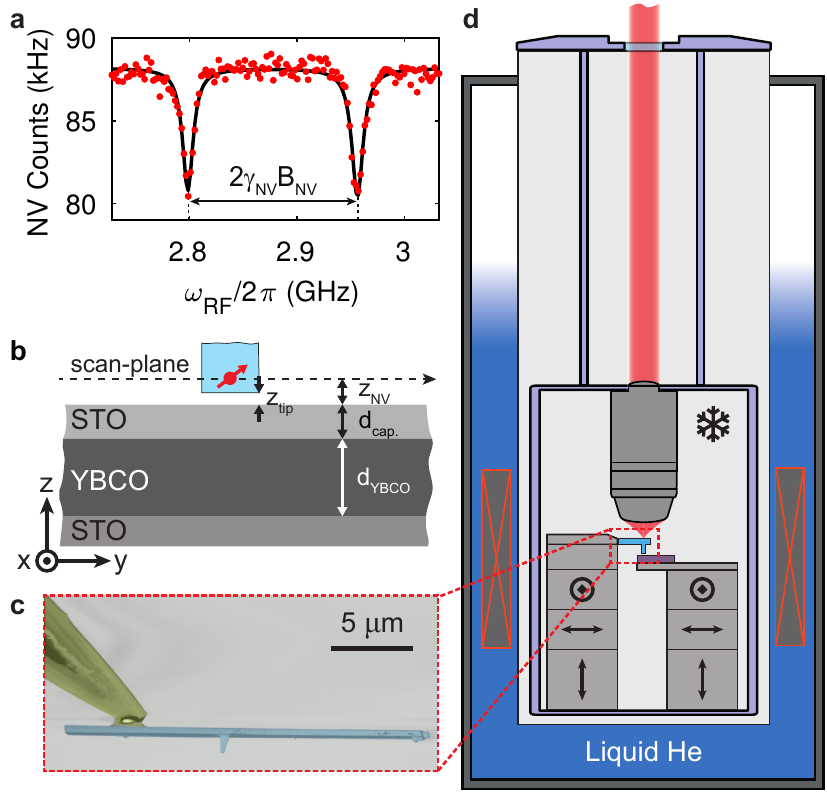}
\caption{Cryogenic scanning NV magnetometer. a) Optically detected magnetic resonance (ODMR) spectrum. The separation between the peaks is proportional to the magnetic field applied to the NV center. b) Geometry of the experiment. A thin film of YBCO was grown on a SrTiO$_3$ (STO) substrate and protected by a capping layer. The red arrow and blue structure indicate the NV spin and diamond nanopillar depicted in c). The distance between the sample surface and the nanopillar (z$_{\textrm{tip}}$) is controlled by means of an atomic-force microscope.
c) False-color electron-microscope image of an all-diamond scanning probe as used in scanning NV magnetometry. The NV sensor spin is located at the apex of the nanopillar visible in the center of the diamond cantilever.
d) Layout of the cryogenic scanning NV magnetometer. Tip and sample scanning is accomplished with three-axis coarse and fine positioning units; NV fluorescence is collected through a tailor-made low-temperature objective. The microscope is kept in a liquid $^4$He bath at a temperature of $4.2~$K.  Figure partially adapted from\,\cite{Thiel2016}.}
\label{FigScanningNVMag}       
\end{figure}

The first realizations of such scanning NV magnetometry employed a diamond nanocrystal attached to an atomic force microscope tip for magnetic imaging\,\cite{Balasubramanian2008,Rondin2011}. However, a significantly more robust and sensitive implementation of scanning NV magnetometry is achieved by using diamond nanopillars containing individual NV centers at their tip, as scanning probes (Fig.\,\ref{FigScanningNVMag}c)\,\cite{Maletinsky2012}. Such scanning probes enabled detecting and imaging the magnetic field generated by a single electron spin\,\cite{Grinolds2013} at imaging resolutions of $\sim20~$nm\,\cite{Maletinsky2012}, which can be further pushed to the sub-nanometer range by employing strong magnetic field gradients\,\cite{Grinolds2014}.

The use of scanning NV centers in all-diamond nanoscale sensors brings several key advantages for sensing and imaging, in particular with respect to their use under cryogenic conditions. Such probes are highly robust and allow for extended measurement times of up to several months. The diamond pillars that constitute a scanning probe act as optical waveguides that render optical excitation and readout of NV centers highly efficient \cite{Babinec2010}. As a result, laser excitation powers of $\sim100~\mu$W are sufficient to yield single-NV count rates of a few $100\,000~$photons/s which, together with typical ODMR contrasts and linewidths as shown in Fig.\,\ref{FigScanningNVMag}a yield sensitivities in the $\mu$T$/\sqrt{\rm{Hz}}$ range - sufficient to detect a (super)current of $100~$nA flowing in a one-dimensional channel at a distance of $10~$nm from the NV center. Additionally, such scanning probes ensure NV-to-sample distances $d_{\rm NV}$ of just a few tens of nm. It is important to note that $d_{\rm NV}$ is the key quantity determining spatial resolution for this imaging method: the amplitude of the magnetic field generated by a source (i.e. spins or current-distributions) modulated at a spatial frequency $1/\lambda$ on a sample surface, decays as $e^{-2\pi d_{\rm NV}/\lambda}$ with $d_{\rm NV}$\,\cite{Roth1989} and such structures therefore become  practically unresolvable for $d_{\rm NV}>\lambda$. Lastly, embedding the NV center in an all-diamond probe for sensing has the advantage of resulting in typically long spin-echo coherence time, on the order of $T_2^{\rm echo}~\sim100~\mu$s, which may be exploited for high-sensitivity ``AC magnetometry'' using coherent manipulation of the NV spin\,\cite{Taylor2008}. For further information regarding such scanning probes and their non-trivial fabrication, we refer the interested reader to recent publications on the topic\,\cite{Maletinsky2012,Appel2016}.

In order to adapt scanning NV magnetometry to cryogenic conditions for the study of nanoscale properties of superconductors, Thiel et al.\,\cite{Thiel2016} constructed a dedicated instrument, which is briefly described here. The setup consists of a liquid-nitrogen-free, $^4$He bath cryostat; a setting which ensures maximal vibrational stability as it avoids excess vibrations originating from liquid-nitrogen boil-off or closed-cycle cryostat operation.
The cryostat is equipped with three superconducting current-driven coils providing an external magnetic field to the sample of up to 0.5 T along each axis.
Vector-field operation allows for aligning magnetic fields to the NV axis. The scanning NV microscope, comprising positioning hardware for the sample and the tip, as well as a microscope objective for optical access, is held in a closed, stainless-steel tube filled with several mbar of $^4$He exchange gas (Fig.\,\ref{FigScanningNVMag}d). The stainless steel tube is directly inserted into the $^4$He bath to ensure efficient cooling of the sample and the scanning probe.

\begin{figure}[t]
\includegraphics[width=\columnwidth]{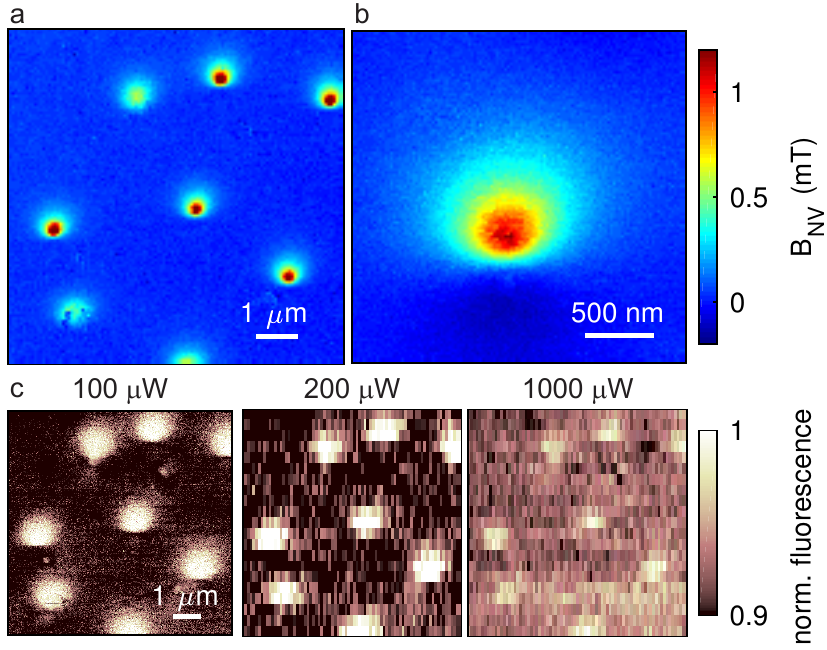}
\caption{Quantitative mapping of stray magnetic fields of vortices in superconducting YBCO, using scanning NV magnetometry. a.) Ensemble and b.) single-vortex image. The stray-field projection onto the NV axis was obtained by measuring Zeeman splitting in the ODMR spectrum at each of the $120\times120$ pixels of the scan. The width of the observed vortex stray field is set by $\Lambda$ (the Pearl length), which is much larger than the estimated spatial resolution of $30~$nm in the image. c.) Iso-B$_{\rm NV}$ images of a vortex arrangement as a function of laser excitation power. The set of data demonstrates non-invasiveness of NV magnetometry in the present experiments: even for the highest powers achievable ($2~$mW), no influence on vortex location could be observed. With increasing laser power, the ODMR contrast decreases as optical pumping leads to relaxation of the electron spin coherences
\cite{Dreau2011}. However, even after the $2~$mW scan, in which no ODMR contrast was observable, the vortices remained unaltered in subsequent images. Figure partially adapted from\,\cite{Thiel2016}.}
\label{FigScanningNVVortexImage}       
\end{figure}

The first applications of cryogenic, nanoscale scanning NV magnetometry focused on imaging of stray magnetic fields emerging from individual vortices in the high-T$_c$ (T$_c\sim90~$K) superconductor YBa$_2$Cu$_3$O$_{7-\delta}$ (YBCO) at temperatures around $4~$K. The study of such vortices was motivated by their fundamental interest\,\cite{Blatter1994}, their relevance for technological applications of SCs (where vortex-motion forms a dominant source of dissipation), and by their comparatively easy to understand stray field, deep in the type-II SC limit (as is the case for YBCO). The experiments were conducted on a $150~$nm thin film of YBCO (Fig.\,\ref{FigScanningNVMag}b), in which vortices were prepared by cooling across $T_c$ in a bias field of $0.4~$mT, applied normal to the sample surface. Stray-field images were subsequently obtained at $4~$K by recording full ODMR traces of the scanning NV center at each pixel comprising the image. Figures\,\ref{FigScanningNVVortexImage}a and b show representative data on a small ensemble of vortices, along with a high-resolution image of a single-vortex stray field, respectively. The images are quantitative in the sense that they present the projection $B_{\rm NV}$ of the vortex stray magnetic field onto the NV axis in physical units, which, for example, allowed the authors to verify that indeed, the imaged vortex carried a quantum of magnetic flux\,\cite{Thiel2016}. This vectorial measurement of $B_{\rm NV}$, along the NV-center axis, is the reason for the non-circularly-symmetric appearance of the vortex stray fields, since the NV axis is tilted from the sample normal by $53^\circ$ (see Fig.\,\ref{FigScanningNVMag}b and discussion below). 

One of the scanning strategies (used, for example, to obtain images in Fig.\,\ref{FigScanningNVVortexImage}c) is the so-called ``iso-B'' imaging, where a single NV is scanned in the presence of constant laser illumination and microwave excitation at a fixed frequency. A reduction in the NV fluorescence is observed whenever the local stray magnetic field fulfills the NV spin-resonance condition. The result is a fast image of vortex location, which, however, can not be quantitatively interpreted in a straightforward way as it does not provide full field information across the whole imaging plane.

The back-action of NV magnetometry on the YBCO sample was tested in this experiment by increasing the readout laser power in steps, up to powers much higher than the values usually used for magnetometry. The essential outcome was that for the present sample and experimental setting, no effect on the vortices was observed. Specifically, Fig.\,\ref{FigScanningNVVortexImage}c shows a series of ``iso-B'' NV-magnetometry images, each taken over the same region of the sample, for increasing laser power up to $1~$mW (as compared to the $100~\mu$W used in Fig.\,\ref{FigScanningNVVortexImage}a and b). The data show no alteration of vortex positions across these scans. While YBCO is known for strong vortex pinning and has a comparatively high T$_c$, prospects for studying more fragile superconductors are excellent, as many opportunities exist to further reduce the back-action of NV magnetometry on the sample with an ultimate total net heat load in the sub-nW range\,\cite{Thiel2016}. Specifically, resonant optical excitation of the NV center\,\cite{Robledo2011} would require no more than a few nW of laser excitation, all-optical spin manipulation\,\cite{Yale2013} could eliminate microwaves and pulsed ESR driving and detection\,\cite{Dreau2011} could lead to laser duty cycles $<1\%$.

\begin{figure}[t]
  \includegraphics[width=\columnwidth]{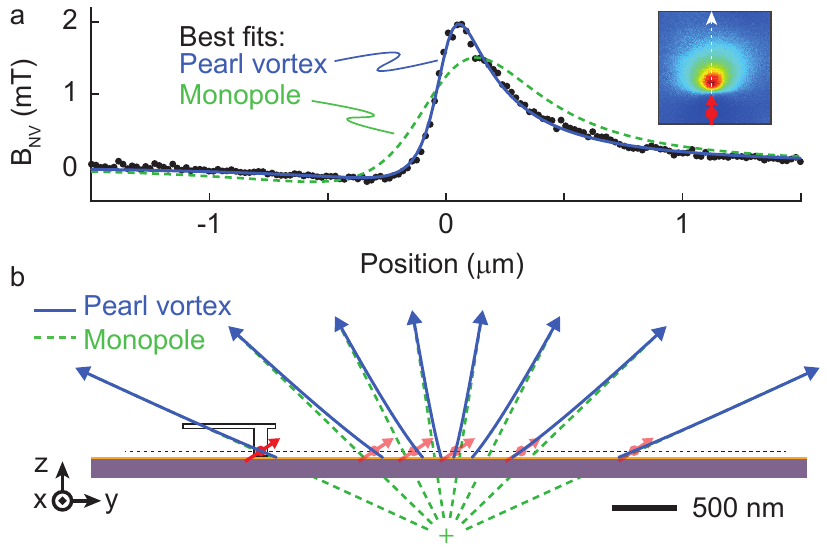}
\caption{Scanning image the vortex field in a YBCO superconductor.  (a) Line-cut through a vortex stray field image (see inset), along with fits to monopole and Pearl models for vortex stray fields (see text). (b) Illustration of the magnetic field lines in the vicinity of a vortex in the monopole and Pearl approximations, together with a schematic representation of the scanning NV center. Figure adapted from\,\cite{Thiel2016}.}
\label{FigQuantVortexImage}       
\end{figure}

The quantitative vortex stray field maps presented in Fig.\,\ref{FigScanningNVVortexImage} allow one to test and compare various models of vortex stray fields and to quantitatively determine the key parameters of the experiment, such as the sample's London penetration depth, $\lambda_{\rm London}$. 
In particular, Thiel et al.\,\cite{Thiel2016} compared the widely used ``monopole approximation''\,\cite{Auslaender2008} to Pearl's analytic solution for vortex stray fields\,\cite{Pearl1964}. 
The former approximates the vortex stray field with one of a monopole and is justified in the far-field of the vortex by the fact that the field-lines generated by the vortex cannot close in the plane of the superconductor by virtue of the Meissner effect. A detailed analysis\,\cite{Pearl1964,Auslaender2008,Carneiro2000} then yields a stray field resembling a virtual magnetic monopole of strength $2\Phi_0$, located a distance $\Lambda=2\lambda_{\rm London}^2/d_{SC}$ below the superconductor surface, where $\Lambda$ is the so-called Pearl length and $d_{SC}$ is the thickness of the superconductor. Given that the strength of this monopole field is set by the flux quantum, the only free parameter in the resulting stray field is the NV-to-monopole distance $z_{\rm NV}+\Lambda$, i.e. changes in $\Lambda$ cannot be distinguished from changes in $z_{\rm NV}$ in a fit to data.
Conversely, the Pearl vortex stray field corresponds to an analytic solution to the second London equation 
\begin{equation}
\nabla\times \vec{j}_s=-\frac{\vec{B}}{\mu_0\lambda_{\rm London}^2}
\end{equation}
in the thin-film limit $d_{SC}\ll\lambda_{\rm London}$, where $\vec{j}_s$ is the superconducting current density, $\vec{B}$ is the total magnetic field and $\mu_0$ is the vacuum permeability. 
This solution yields the supercurrent distribution around the vortex and, by virtue of Biot-Savard's law, the vortex magnetic stray field outside of the superconductor. 
The comparison of the two quantitative fits in Fig.\,\ref{FigQuantVortexImage}a clearly shows the superiority of the Pearl model over the monopole approximation. Past works on local imaging of vortex stray fields successfully employed the monopole model\,\cite{Auslaender2008}, since monopole and Pearl models converge to essentially the same solution for probe-sample distances $>\lambda_{\rm London}$. In closer proximity to the sample, however, this is not the case and strong deviations of the predicted field lines exist for the two cases (Fig.\,\ref{FigQuantVortexImage}b). The Pearl model has only two independent free parameters, $d_{\rm NV}$ and the Pearl length $\Lambda=2\lambda_{\rm London}^2/d_{SC}$. For these, the fit in Fig.\,\ref{FigQuantVortexImage}a resulted in $\Lambda=840\pm20~$nm and $d_{\rm NV}=104\pm2~$nm, where the error was extracted from the statistical uncertainty of the fit. Note that $d_{\rm NV}$ here refers to the distance between the NV center and the center of the superconducting layer; taking into account the value of $d_{SC}$ as well as the capping layer protecting the superconductor yielded the distance between NV and sample surface and therefore the spatial resolution of this imaging method. 
Note that a recent, unpublished analysis by the authors of\,\cite{Thiel2016} indicated that treating the SC sample as a layer of nonzero thickness (i.e. accounting for the fact that $d_{SC}\sim\lambda_{\rm London}$) yields a slight correction for $d_{\rm NV}$ and results in a final NV-to-sample distance, i.e. spatial resolution, of $\sim30~$nm for this experiment. 
For this refined analysis, the authors employed the full analytic solution to the second London equation for arbitrary sample thicknesses as derived by Carneiro et al.\,\cite{Carneiro2000} to determine $\vec{j}_s$ and subsequently determined the vortex stray field through Biot-Savard's law. 

Another experiment has recently also demonstrated imaging of vortex stray fields using scanning NV magnetometry\,\cite{Pelliccione2016}. In contrast to the work of Ref.\,\cite{Thiel2016}  (presented in Fig.\,\ref{FigScanningNVVortexImage} and Fig.\,\ref{FigQuantVortexImage}), the authors of\,\cite{Pelliccione2016} employed a closed-cycle cryostat - an interesting experimental variant, which improves usability and simplicity of the setup at the price of reduced available magnetic field strengths and increased vibrations. Using their setup, these authors employed iso-magnetic field imaging for locating individual vortices. The results of\,\cite{Pelliccione2016} confirm the utility of NV magnetometry for the study  of superconductors at the nanoscale.

These advances in NV magnetometry of vortex stray fields offer attractive prospects for future nanoscale explorations of superconductors and at the same time still have significant potential for further improvements in imaging and sensing performance.

The key sensing parameters indicated above, the $\approx\,30\,$nm spatial resolution and imaging times of a few hours per frame at the SNR and sampling rate indicated, are limited by both fundamental and technical factors. Spatial resolution is set by the NV-sample distance $d_{\rm NV}$, which is determined by the depth of the NV within the scanning pillar, the AFM tip-sample distance control and potential contaminants on the sample surface. We estimate that for the present experiments the first factor dominates as $d_{\rm NV}$ agrees well with the expected implantation depth for the nitrogen ions used to generate NV centers in the diamond tips\,\cite{Thiel2016,Pelliccione2016,Pezzagna2010}. The implantation depth can be further reduced to the range of a few nm\,\cite{Pezzagna2010}, which would bring a corresponding improvement in spatial resolution, but likely also a reduction in NV coherence times\,\cite{Myers2014} and therefore sensitivity. Creating very shallow NV centers, which at the same time allow for a several-nm resolution and high sensitivity is subject to ongoing research\,\cite{Lesik2016}. Sensitivity and therefore data acquisition time are further limited by the detected photon count rates from the NV center and the ODMR readout contrast. While the latter has been largely optimised to $\sim20\%$ in the present devices, photon collection efficiencies can still be significantly increased boosting the $\sim100\,000~$counts/s rates achieved in\,\cite{Thiel2016}. In fact, recent unpublished results\,\cite{Shields2018} have demonstrated photon count rates approaching $\sim2\,000\,000~$counts/s, which will yield a $20-$fold improvement in the data-acquisition time over currently published results.

Arguably the most successful and advanced technologies to locally address superconductors on the nanoscale thus far have been scanning tunneling microscopy (STM)\,\cite{Fischer2007} and scanning superconducting interference device (SQUID)\,\cite{Kirtley2010} microscopy. STM has the unmatched advantages of spatial resolution and ability to probe the SC state at the upper critical field~\cite{Suderow2014}.  Scanning SQUIDs have recently made remarkable advances with implementations of nanoscale SQUID-on-tip architectures\,\cite{Vasyukov2013,Embon2015,Embon2017} achieving magnetic imaging at spatial resolutions in the $50~$nm range. While sub-nm resolution of STM and the exquisite sensitivity of SQUIDs are currently out reach for NV magnetometry, it is the combined sensing performance and the wide range of operating conditions for NV magnetometry, which brings attractive prospects for future applications of this technique, in particular to superconductivity. 
\\
\\
\\
\\

\section{Conclusions and outlook}
\label{dec:Conclusion}

NV magnetometry can operate over the whole relevant temperature range from cryogenic conditions to room temperature. It can, in principle, operate at arbitrarily high magnetic fields (as long as these are aligned with the NV axis) and it can be applied to partially or fully insulating samples. Furthermore, NV-based sensing can address a remarkable range of frequencies in magnetometry, from DC (using ODMR as shown above\,\cite{Rondin2014}) and over MHz (using ``dynamical decoupling'' based sensing approaches\,\cite{Degen2017}) all the way to the GHz range (where ``spin relaxometry'' can be employed for sensing\,\cite{Casola2018}). 
The advantages of NV-based magnetometers (e.g., the ability to perform quantitative vectorial measurements in a wide temperature range) hold promise for these devices in exploring several superconducting phenomena of current interest such as island superconductivity \cite{Kresin2006,Wolf2012}, surface superconducting states \cite{Zeinali2016}, and for measuring the pinning strength \cite{Embon2015} and dynamics of vortices \cite{vortexdynamicszeldov2017}. 

As the NV-based methods develop further, one day it may become possible  to image the spin-fabric in cuprate superconductors \cite{Luscher2007}; the required spatial resolution is $\approx\,0.3\,$nm. It would be interesting to also apply NV-center based noise spectroscopy \cite{Kolkowitz2015} to the study of superconductors. 

One interesting research direction is exploration of superconductivity in diamond itself. Indeed, diamond is known to become superconducting at sufficient levels of doping with boron (see, for example, \cite{Kumar2017} and references therein). In this system, one can imagine the superconductor under study and its probe being one and the same material, perhaps, differentially implanted/doped to optimize the experimental geometry. 

Another nascent direction is using NV magnetometry to study superconducting phenomena at high pressure \cite{Doherty2014Pressure}. Here the NV centers can either be incorporated into the diamond anvils used to produce high pressure or used as magnetic probes ``strategically'' placed in the vicinity of the sample under study. The NV-bearing diamond could be in the form of a plate or nanodiamond and can be either under high pressure or not depending on the experimental design. Perhaps, NV-based methods can make a contribution in the diagnostics of SC materials at high pressure, where the record value of T$_{\rm{c}}=203$\,K at a pressure around 150\,GPa has been observed in sulfur hydrides \cite{Drozdov2015}
and where further advances towards room-temperature superconductivity are anticipated \cite{Gorkov2018}.

There appears to be considerable interest on the part of the superconductivity-research community in utilizing the capabilities offered by both the scanning and wide-field-imaging modalities described in this review, so  further rapid development of techniques and applications can be expected. We would like to encourage expanding the productive dialogue between the superconductivity and NV-magnetometry communities. 

\begin{acknowledgements}
The authors thank V.\,Z.\,Kresin for initiating and encouraging this review and M.\,L.\,Cohen, M.\,Eremets, J.-F.\,Roche, O.\,P.\,Sushkov, and N.\,Yao for useful discussions. 
We gratefully acknowledge financial support through the NCCR QSIT, a competence center funded by the Swiss NSF, through the Swiss Nanoscience Institute, by the EU FP7 project DIADEMS (grant \#611143), 
and through SNF Project Grant No. 169321 and No. 155845; this work was additionally supported by the German Federal Ministry of Education and Research (BMBF) within the Quantumtechnologien program (FKZ 13N14439), the DFG DIP project Ref. FO 703/2-1 and by the Israeli Science Foundation. V.\,M.\,Acosta acknowledges funding support by the Beckman Young Investigator Program.

\end{acknowledgements}


\begin{thebibliography}{82}%
\makeatletter
\providecommand \@ifxundefined [1]{%
 \@ifx{#1\undefined}
}%
\providecommand \@ifnum [1]{%
 \ifnum #1\expandafter \@firstoftwo
 \else \expandafter \@secondoftwo
 \fi
}%
\providecommand \@ifx [1]{%
 \ifx #1\expandafter \@firstoftwo
 \else \expandafter \@secondoftwo
 \fi
}%
\providecommand \natexlab [1]{#1}%
\providecommand \enquote  [1]{``#1''}%
\providecommand \bibnamefont  [1]{#1}%
\providecommand \bibfnamefont [1]{#1}%
\providecommand \citenamefont [1]{#1}%
\providecommand \href@noop [0]{\@secondoftwo}%
\providecommand \href [0]{\begingroup \@sanitize@url \@href}%
\providecommand \@href[1]{\@@startlink{#1}\@@href}%
\providecommand \@@href[1]{\endgroup#1\@@endlink}%
\providecommand \@sanitize@url [0]{\catcode `\\12\catcode `\$12\catcode
  `\&12\catcode `\#12\catcode `\^12\catcode `\_12\catcode `\%12\relax}%
\providecommand \@@startlink[1]{}%
\providecommand \@@endlink[0]{}%
\providecommand \url  [0]{\begingroup\@sanitize@url \@url }%
\providecommand \@url [1]{\endgroup\@href {#1}{\urlprefix }}%
\providecommand \urlprefix  [0]{URL }%
\providecommand \Eprint [0]{\href }%
\providecommand \doibase [0]{http://dx.doi.org/}%
\providecommand \selectlanguage [0]{\@gobble}%
\providecommand \bibinfo  [0]{\@secondoftwo}%
\providecommand \bibfield  [0]{\@secondoftwo}%
\providecommand \translation [1]{[#1]}%
\providecommand \BibitemOpen [0]{}%
\providecommand \bibitemStop [0]{}%
\providecommand \bibitemNoStop [0]{.\EOS\space}%
\providecommand \EOS [0]{\spacefactor3000\relax}%
\providecommand \BibitemShut  [1]{\csname bibitem#1\endcsname}%
\let\auto@bib@innerbib\@empty
\bibitem [{\citenamefont {Bending}(1999)}]{Bending1999}%
  \BibitemOpen
  \bibfield  {author} {\bibinfo {author} {\bibfnamefont {S.~J.}\ \bibnamefont
  {Bending}},\ }\href {https://doi.org/10.1080/000187399243437} {\bibfield
  {journal} {\bibinfo  {journal} {Advances in Physics}\ }\textbf {\bibinfo
  {volume} {48}},\ \bibinfo {pages} {449} (\bibinfo {year} {1999})}\BibitemShut
  {NoStop}%
\bibitem [{\citenamefont {Kresin}\ \emph {et~al.}(2006)\citenamefont {Kresin},
  \citenamefont {Ovchinnikov},\ and\ \citenamefont {Wolf}}]{Kresin2006}%
  \BibitemOpen
  \bibfield  {author} {\bibinfo {author} {\bibfnamefont {V.~Z.}\ \bibnamefont
  {Kresin}}, \bibinfo {author} {\bibfnamefont {Y.~N.}\ \bibnamefont
  {Ovchinnikov}}, \ and\ \bibinfo {author} {\bibfnamefont {S.~A.}\ \bibnamefont
  {Wolf}},\ }\href {http://dx.doi.org/10.1016/j.physrep.2006.05.006} {\bibfield
   {journal} {\bibinfo  {journal} {Physics Reports}\ }\textbf {\bibinfo
  {volume} {431}},\ \bibinfo {pages} {231} (\bibinfo {year}
  {2006})}\BibitemShut {NoStop}%
\bibitem [{\citenamefont {Wolf}\ and\ \citenamefont {Kresin}(2012)}]{Wolf2012}%
  \BibitemOpen
  \bibfield  {author} {\bibinfo {author} {\bibfnamefont {S.~A.}\ \bibnamefont
  {Wolf}}\ and\ \bibinfo {author} {\bibfnamefont {V.~Z.}\ \bibnamefont
  {Kresin}},\ }\href@noop {} {\emph {\bibinfo {title} {Novel
  superconductivity}}}\ (\bibinfo  {publisher} {Springer Science \& Business
  Media},\ \bibinfo {year} {2012})\BibitemShut {NoStop}%
\bibitem [{\citenamefont {Kresin}\ and\ \citenamefont
  {Ovchinnikov}(2008)}]{Kresin2008}%
  \BibitemOpen
  \bibfield  {author} {\bibinfo {author} {\bibfnamefont {V.~Z.}\ \bibnamefont
  {Kresin}}\ and\ \bibinfo {author} {\bibfnamefont {Y.~N.}\ \bibnamefont
  {Ovchinnikov}},\ }\href {http://stacks.iop.org/1063-7869/51/i=5/a=R01}
  {\bibfield  {journal} {\bibinfo  {journal} {Physics-Uspekhi}\ }\textbf
  {\bibinfo {volume} {51}},\ \bibinfo {pages} {427} (\bibinfo {year}
  {2008})}\BibitemShut {NoStop}%
\bibitem [{\citenamefont {Kresin}(2012)}]{Kresin2012}%
  \BibitemOpen
  \bibfield  {author} {\bibinfo {author} {\bibfnamefont {V.}~\bibnamefont
  {Kresin}},\ }\href {\doibase 10.1007/s10948-012-1439-y} {\bibfield  {journal}
  {\bibinfo  {journal} {Journal of Superconductivity and Novel Magnetism}\
  }\textbf {\bibinfo {volume} {25}},\ \bibinfo {pages} {711} (\bibinfo {year}
  {2012})}\BibitemShut {NoStop}%
\bibitem [{\citenamefont {Kresin}\ and\ \citenamefont
  {Ovchinnikov}(2013)}]{Kresin2013}%
  \BibitemOpen
  \bibfield  {author} {\bibinfo {author} {\bibfnamefont {V.~Z.}\ \bibnamefont
  {Kresin}}\ and\ \bibinfo {author} {\bibfnamefont {Y.~N.}\ \bibnamefont
  {Ovchinnikov}},\ }\href {\doibase 10.1007/s10948-012-1961-y} {\bibfield
  {journal} {\bibinfo  {journal} {Journal of Superconductivity and Novel
  Magnetism}\ }\textbf {\bibinfo {volume} {26}},\ \bibinfo {pages} {745}
  (\bibinfo {year} {2013})}\BibitemShut {NoStop}%
\bibitem [{\citenamefont {Ams{\"u}ss}\ \emph {et~al.}(2014)\citenamefont
  {Ams{\"u}ss}, \citenamefont {Saito},\ and\ \citenamefont
  {Munro}}]{AMSUSS2014}%
  \BibitemOpen
  \bibfield  {author} {\bibinfo {author} {\bibfnamefont {R.}~\bibnamefont
  {Ams{\"u}ss}}, \bibinfo {author} {\bibfnamefont {S.}~\bibnamefont {Saito}}, \
  and\ \bibinfo {author} {\bibfnamefont {W.}~\bibnamefont {Munro}},\ }in\ \href
  {\doibase https://doi.org/10.1533/9780857096685.2.264} {\emph {\bibinfo
  {booktitle} {Quantum Information Processing with Diamond}}},\ \bibinfo
  {editor} {edited by\ \bibinfo {editor} {\bibfnamefont {S.}~\bibnamefont
  {Prawer}}\ and\ \bibinfo {editor} {\bibfnamefont {I.}~\bibnamefont
  {Aharonovich}}}\ (\bibinfo  {publisher} {Woodhead Publishing},\ \bibinfo
  {year} {2014})\ pp.\ \bibinfo {pages} {264 -- 290}\BibitemShut {NoStop}%
\bibitem [{\citenamefont {Chernobrod}\ and\ \citenamefont
  {Berman}(2005)}]{Chernobrod2005}%
  \BibitemOpen
  \bibfield  {author} {\bibinfo {author} {\bibfnamefont {B.~M.}\ \bibnamefont
  {Chernobrod}}\ and\ \bibinfo {author} {\bibfnamefont {G.~P.}\ \bibnamefont
  {Berman}},\ }\href {https://doi.org/10.1063/1.1829373} {\bibfield  {journal}
  {\bibinfo  {journal} {Journal of Applied Physics}\ }\textbf {\bibinfo
  {volume} {97}},\ \bibinfo {pages} {014903} (\bibinfo {year}
  {2005})}\BibitemShut {NoStop}%
\bibitem [{\citenamefont {Taylor}\ \emph {et~al.}(2008)\citenamefont {Taylor},
  \citenamefont {Cappellaro}, \citenamefont {Childress}, \citenamefont {Jiang},
  \citenamefont {Neumann}, \citenamefont {Budker}, \citenamefont {Hemmer},
  \citenamefont {Yacoby}, \citenamefont {Walsworth},\ and\ \citenamefont
  {Lukin}}]{Taylor2008}%
  \BibitemOpen
  \bibfield  {author} {\bibinfo {author} {\bibfnamefont {J.~M.}\ \bibnamefont
  {Taylor}}, \bibinfo {author} {\bibfnamefont {P.}~\bibnamefont {Cappellaro}},
  \bibinfo {author} {\bibfnamefont {L.}~\bibnamefont {Childress}}, \bibinfo
  {author} {\bibfnamefont {L.}~\bibnamefont {Jiang}}, \bibinfo {author}
  {\bibfnamefont {P.}~\bibnamefont {Neumann}}, \bibinfo {author} {\bibfnamefont
  {D.}~\bibnamefont {Budker}}, \bibinfo {author} {\bibfnamefont {P.~R.}\
  \bibnamefont {Hemmer}}, \bibinfo {author} {\bibfnamefont {A.}~\bibnamefont
  {Yacoby}}, \bibinfo {author} {\bibfnamefont {R.}~\bibnamefont {Walsworth}}, \
  and\ \bibinfo {author} {\bibfnamefont {M.~D.}\ \bibnamefont {Lukin}},\ }\href
  {\doibase 10.1038/nphys1075} {\bibfield  {journal} {\bibinfo  {journal}
  {Nature Physics}\ }\textbf {\bibinfo {volume} {4}},\ \bibinfo {pages} {810}
  (\bibinfo {year} {2008})}\BibitemShut {NoStop}%
\bibitem [{\citenamefont {Degen}(2008)}]{Degen2008}%
  \BibitemOpen
  \bibfield  {author} {\bibinfo {author} {\bibfnamefont {C.~L.}\ \bibnamefont
  {Degen}},\ }\href {https://doi.org/10.1063/1.2943282} {\bibfield  {journal}
  {\bibinfo  {journal} {Applied Physics Letters}\ }\textbf {\bibinfo {volume}
  {92}} (\bibinfo {year} {2008})}\BibitemShut {NoStop}%
\bibitem [{\citenamefont {Acosta}\ \emph
  {et~al.}(2010{\natexlab{a}})\citenamefont {Acosta}, \citenamefont {Bauch},
  \citenamefont {Ledbetter}, \citenamefont {Waxman}, \citenamefont {Bouchard},\
  and\ \citenamefont {Budker}}]{TempBudker}%
  \BibitemOpen
  \bibfield  {author} {\bibinfo {author} {\bibfnamefont {V.~M.}\ \bibnamefont
  {Acosta}}, \bibinfo {author} {\bibfnamefont {E.}~\bibnamefont {Bauch}},
  \bibinfo {author} {\bibfnamefont {M.~P.}\ \bibnamefont {Ledbetter}}, \bibinfo
  {author} {\bibfnamefont {A.}~\bibnamefont {Waxman}}, \bibinfo {author}
  {\bibfnamefont {L.-S.}\ \bibnamefont {Bouchard}}, \ and\ \bibinfo {author}
  {\bibfnamefont {D.}~\bibnamefont {Budker}},\ }\href {\doibase
  10.1103/PhysRevLett.104.070801} {\bibfield  {journal} {\bibinfo  {journal}
  {Phys. Rev. Lett.}\ }\textbf {\bibinfo {volume} {104}},\ \bibinfo {pages}
  {070801} (\bibinfo {year} {2010}{\natexlab{a}})}\BibitemShut {NoStop}%
\bibitem [{\citenamefont {Acosta}\ \emph
  {et~al.}(2010{\natexlab{b}})\citenamefont {Acosta}, \citenamefont {Bauch},
  \citenamefont {Jarmola}, \citenamefont {Zipp}, \citenamefont {Ledbetter},\
  and\ \citenamefont {Budker}}]{Acosta2010a}%
  \BibitemOpen
  \bibfield  {author} {\bibinfo {author} {\bibfnamefont {V.~M.}\ \bibnamefont
  {Acosta}}, \bibinfo {author} {\bibfnamefont {E.}~\bibnamefont {Bauch}},
  \bibinfo {author} {\bibfnamefont {A.}~\bibnamefont {Jarmola}}, \bibinfo
  {author} {\bibfnamefont {L.~J.}\ \bibnamefont {Zipp}}, \bibinfo {author}
  {\bibfnamefont {M.~P.}\ \bibnamefont {Ledbetter}}, \ and\ \bibinfo {author}
  {\bibfnamefont {D.}~\bibnamefont {Budker}},\ }\href
  {https://doi.org/10.1063/1.3507884} {\bibfield  {journal} {\bibinfo
  {journal} {Applied Physics Letters}\ }\textbf {\bibinfo {volume} {97}},\
  \bibinfo {eid} {174104} (\bibinfo {year} {2010}{\natexlab{b}})}\BibitemShut
  {NoStop}%
\bibitem [{\citenamefont {Dolde}\ \emph {et~al.}(2011)\citenamefont {Dolde},
  \citenamefont {Fedder}, \citenamefont {Doherty}, \citenamefont {N\"obauer},
  \citenamefont {Rempp}, \citenamefont {Balasubramanian}, \citenamefont {Wolf},
  \citenamefont {Reinhard}, \citenamefont {Hollenberg}, \citenamefont
  {Jelezko},\ and\ \citenamefont {Wrachtrup}}]{Dolde2011}%
  \BibitemOpen
  \bibfield  {author} {\bibinfo {author} {\bibfnamefont {F.}~\bibnamefont
  {Dolde}}, \bibinfo {author} {\bibfnamefont {H.}~\bibnamefont {Fedder}},
  \bibinfo {author} {\bibfnamefont {M.~W.}\ \bibnamefont {Doherty}}, \bibinfo
  {author} {\bibfnamefont {T.}~\bibnamefont {N\"obauer}}, \bibinfo {author}
  {\bibfnamefont {F.}~\bibnamefont {Rempp}}, \bibinfo {author} {\bibfnamefont
  {G.}~\bibnamefont {Balasubramanian}}, \bibinfo {author} {\bibfnamefont
  {T.}~\bibnamefont {Wolf}}, \bibinfo {author} {\bibfnamefont {F.}~\bibnamefont
  {Reinhard}}, \bibinfo {author} {\bibfnamefont {L.~C.~L.}\ \bibnamefont
  {Hollenberg}}, \bibinfo {author} {\bibfnamefont {F.}~\bibnamefont {Jelezko}},
  \ and\ \bibinfo {author} {\bibfnamefont {J.}~\bibnamefont {Wrachtrup}},\
  }\href {\doibase 10.1038/nphys1969} {\bibfield  {journal} {\bibinfo
  {journal} {Nature Physics}\ }\textbf {\bibinfo {volume} {7}},\ \bibinfo
  {pages} {459} (\bibinfo {year} {2011})}\BibitemShut {NoStop}%
\bibitem [{\citenamefont {Rondin}\ \emph {et~al.}(2014)\citenamefont {Rondin},
  \citenamefont {Tetienne}, \citenamefont {Hingant}, \citenamefont {Roch},
  \citenamefont {Maletinsky},\ and\ \citenamefont {Jacques}}]{Rondin2014}%
  \BibitemOpen
  \bibfield  {author} {\bibinfo {author} {\bibfnamefont {L.}~\bibnamefont
  {Rondin}}, \bibinfo {author} {\bibfnamefont {J.-P.}\ \bibnamefont
  {Tetienne}}, \bibinfo {author} {\bibfnamefont {T.}~\bibnamefont {Hingant}},
  \bibinfo {author} {\bibfnamefont {J.-F.}\ \bibnamefont {Roch}}, \bibinfo
  {author} {\bibfnamefont {P.}~\bibnamefont {Maletinsky}}, \ and\ \bibinfo
  {author} {\bibfnamefont {V.}~\bibnamefont {Jacques}},\ }\href
  {https://doi.org/10.1088/0034-4885/77/5/056503} {\bibfield  {journal}
  {\bibinfo  {journal} {Reports on Progress in Physics}\ }\textbf {\bibinfo
  {volume} {77}},\ \bibinfo {pages} {056503} (\bibinfo {year}
  {2014})}\BibitemShut {NoStop}%
\bibitem [{\citenamefont {Jelezko}\ and\ \citenamefont
  {Wrachtrup}(2006)}]{Jelezko2006a}%
  \BibitemOpen
  \bibfield  {author} {\bibinfo {author} {\bibfnamefont {F.}~\bibnamefont
  {Jelezko}}\ and\ \bibinfo {author} {\bibfnamefont {J.}~\bibnamefont
  {Wrachtrup}},\ }\href {\doibase 10.1002/pssa.200671403} {\bibfield  {journal}
  {\bibinfo  {journal} {physica status solidi (a)}\ }\textbf {\bibinfo {volume}
  {203}},\ \bibinfo {pages} {3207} (\bibinfo {year} {2006})}\BibitemShut
  {NoStop}%
\bibitem [{\citenamefont {Robledo}\ \emph
  {et~al.}(2011{\natexlab{a}})\citenamefont {Robledo}, \citenamefont {Bernien},
  \citenamefont {van~der Sar},\ and\ \citenamefont {Hanson}}]{Robledo2011NJP}%
  \BibitemOpen
  \bibfield  {author} {\bibinfo {author} {\bibfnamefont {L.}~\bibnamefont
  {Robledo}}, \bibinfo {author} {\bibfnamefont {H.}~\bibnamefont {Bernien}},
  \bibinfo {author} {\bibfnamefont {T.}~\bibnamefont {van~der Sar}}, \ and\
  \bibinfo {author} {\bibfnamefont {R.}~\bibnamefont {Hanson}},\ }\href
  {http://stacks.iop.org/1367-2630/13/i=2/a=025013} {\bibfield  {journal}
  {\bibinfo  {journal} {New Journal of Physics}\ }\textbf {\bibinfo {volume}
  {13}},\ \bibinfo {pages} {025013} (\bibinfo {year}
  {2011}{\natexlab{a}})}\BibitemShut {NoStop}%
\bibitem [{\citenamefont {Dumeige}\ \emph {et~al.}(2013)\citenamefont
  {Dumeige}, \citenamefont {Chipaux}, \citenamefont {Jacques}, \citenamefont
  {Treussart}, \citenamefont {Roch}, \citenamefont {Debuisschert},
  \citenamefont {Acosta}, \citenamefont {Jarmola}, \citenamefont {Jensen},
  \citenamefont {Kehayias},\ and\ \citenamefont {Budker}}]{Dumeige2013}%
  \BibitemOpen
  \bibfield  {author} {\bibinfo {author} {\bibfnamefont {Y.}~\bibnamefont
  {Dumeige}}, \bibinfo {author} {\bibfnamefont {M.}~\bibnamefont {Chipaux}},
  \bibinfo {author} {\bibfnamefont {V.}~\bibnamefont {Jacques}}, \bibinfo
  {author} {\bibfnamefont {F.}~\bibnamefont {Treussart}}, \bibinfo {author}
  {\bibfnamefont {J.-F.}\ \bibnamefont {Roch}}, \bibinfo {author}
  {\bibfnamefont {T.}~\bibnamefont {Debuisschert}}, \bibinfo {author}
  {\bibfnamefont {V.}~\bibnamefont {Acosta}}, \bibinfo {author} {\bibfnamefont
  {A.}~\bibnamefont {Jarmola}}, \bibinfo {author} {\bibfnamefont
  {K.}~\bibnamefont {Jensen}}, \bibinfo {author} {\bibfnamefont
  {P.}~\bibnamefont {Kehayias}}, \ and\ \bibinfo {author} {\bibfnamefont
  {D.}~\bibnamefont {Budker}},\ }\href
  {https://doi.org/10.1103/PhysRevB.87.155202} {\bibfield  {journal} {\bibinfo
  {journal} {Physical Review B}\ }\textbf {\bibinfo {volume} {87}} (\bibinfo
  {year} {2013})}\BibitemShut {NoStop}%
\bibitem [{\citenamefont {Kalb}\ \emph {et~al.}(2018)\citenamefont {Kalb},
  \citenamefont {Humphreys}, \citenamefont {Slim},\ and\ \citenamefont
  {Hanson}}]{Kalb2018}%
  \BibitemOpen
  \bibfield  {author} {\bibinfo {author} {\bibfnamefont {N.}~\bibnamefont
  {Kalb}}, \bibinfo {author} {\bibfnamefont {P.~C.}\ \bibnamefont {Humphreys}},
  \bibinfo {author} {\bibfnamefont {J.~J.}\ \bibnamefont {Slim}}, \ and\
  \bibinfo {author} {\bibfnamefont {R.}~\bibnamefont {Hanson}},\ }\href
  {\doibase 10.1103/PhysRevA.97.062330} {\bibfield  {journal} {\bibinfo
  {journal} {Phys. Rev. A}\ }\textbf {\bibinfo {volume} {97}},\ \bibinfo
  {pages} {062330} (\bibinfo {year} {2018})}\BibitemShut {NoStop}%
\bibitem [{\citenamefont {Tetienne}\ \emph {et~al.}(2012)\citenamefont
  {Tetienne}, \citenamefont {Rondin}, \citenamefont {Spinicelli}, \citenamefont
  {Chipaux}, \citenamefont {Debuisschert}, \citenamefont {Roch},\ and\
  \citenamefont {Jacques}}]{Tetienne2012}%
  \BibitemOpen
  \bibfield  {author} {\bibinfo {author} {\bibfnamefont {J.}~\bibnamefont
  {Tetienne}}, \bibinfo {author} {\bibfnamefont {L.}~\bibnamefont {Rondin}},
  \bibinfo {author} {\bibfnamefont {P.}~\bibnamefont {Spinicelli}}, \bibinfo
  {author} {\bibfnamefont {M.}~\bibnamefont {Chipaux}}, \bibinfo {author}
  {\bibfnamefont {T.}~\bibnamefont {Debuisschert}}, \bibinfo {author}
  {\bibfnamefont {J.}~\bibnamefont {Roch}}, \ and\ \bibinfo {author}
  {\bibfnamefont {V.}~\bibnamefont {Jacques}},\ }\href
  {http://dx.doi.org/10.1088/1367-2630/14/10/103033} {\bibfield  {journal}
  {\bibinfo  {journal} {New Journal of Physics}\ }\textbf {\bibinfo {volume}
  {14}},\ \bibinfo {pages} {103033} (\bibinfo {year} {2012})}\BibitemShut
  {NoStop}%
\bibitem [{\citenamefont {Doherty}\ \emph {et~al.}(2013)\citenamefont
  {Doherty}, \citenamefont {Manson}, \citenamefont {Delaney}, \citenamefont
  {Jelezko}, \citenamefont {Wrachtrup},\ and\ \citenamefont
  {Hollenberg}}]{Doherty2013}%
  \BibitemOpen
  \bibfield  {author} {\bibinfo {author} {\bibfnamefont {M.~W.}\ \bibnamefont
  {Doherty}}, \bibinfo {author} {\bibfnamefont {N.~B.}\ \bibnamefont {Manson}},
  \bibinfo {author} {\bibfnamefont {P.}~\bibnamefont {Delaney}}, \bibinfo
  {author} {\bibfnamefont {F.}~\bibnamefont {Jelezko}}, \bibinfo {author}
  {\bibfnamefont {J.}~\bibnamefont {Wrachtrup}}, \ and\ \bibinfo {author}
  {\bibfnamefont {L.~C.}\ \bibnamefont {Hollenberg}},\ }\href
  {http://dx.doi.org/10.1016/j.physrep.2013.02.001} {\bibfield  {journal}
  {\bibinfo  {journal} {Physics Reports}\ }\textbf {\bibinfo {volume} {528}},\
  \bibinfo {pages} {1} (\bibinfo {year} {2013})}\BibitemShut {NoStop}%
\bibitem [{\citenamefont {Jensen}\ \emph {et~al.}(2017)\citenamefont {Jensen},
  \citenamefont {Kehayias},\ and\ \citenamefont {Budker}}]{Jensen2017}%
  \BibitemOpen
  \bibfield  {author} {\bibinfo {author} {\bibfnamefont {K.}~\bibnamefont
  {Jensen}}, \bibinfo {author} {\bibfnamefont {P.}~\bibnamefont {Kehayias}}, \
  and\ \bibinfo {author} {\bibfnamefont {D.}~\bibnamefont {Budker}},\ }\enquote
  {\bibinfo {title} {High sensitivity magnetometers},}\ \ (\bibinfo
  {publisher} {Springer International Publishing},\ \bibinfo {year} {2017})\
  Chap.\ \bibinfo {chapter} {Magnetometry with Nitrogen-Vacancy Centers in
  Diamond}, pp.\ \bibinfo {pages} {553--576}\BibitemShut {NoStop}%
\bibitem [{\citenamefont {Siyushev}\ \emph {et~al.}(2010)\citenamefont
  {Siyushev}, \citenamefont {Kaiser}, \citenamefont {Jacques}, \citenamefont
  {Gerhardt}, \citenamefont {Bischof}, \citenamefont {Fedder}, \citenamefont
  {Dodson}, \citenamefont {Markham}, \citenamefont {Twitchen}, \citenamefont
  {Jelezko},\ and\ \citenamefont {Wrachtrup}}]{Siyushev2010}%
  \BibitemOpen
  \bibfield  {author} {\bibinfo {author} {\bibfnamefont {P.}~\bibnamefont
  {Siyushev}}, \bibinfo {author} {\bibfnamefont {F.}~\bibnamefont {Kaiser}},
  \bibinfo {author} {\bibfnamefont {V.}~\bibnamefont {Jacques}}, \bibinfo
  {author} {\bibfnamefont {I.}~\bibnamefont {Gerhardt}}, \bibinfo {author}
  {\bibfnamefont {S.}~\bibnamefont {Bischof}}, \bibinfo {author} {\bibfnamefont
  {H.}~\bibnamefont {Fedder}}, \bibinfo {author} {\bibfnamefont
  {J.}~\bibnamefont {Dodson}}, \bibinfo {author} {\bibfnamefont
  {M.}~\bibnamefont {Markham}}, \bibinfo {author} {\bibfnamefont
  {D.}~\bibnamefont {Twitchen}}, \bibinfo {author} {\bibfnamefont
  {F.}~\bibnamefont {Jelezko}}, \ and\ \bibinfo {author} {\bibfnamefont
  {J.}~\bibnamefont {Wrachtrup}},\ }\href {\doibase 10.1063/1.3519849}
  {\bibfield  {journal} {\bibinfo  {journal} {Applied Physics Letters}\
  }\textbf {\bibinfo {volume} {97}},\ \bibinfo {pages} {241902} (\bibinfo
  {year} {2010})}\BibitemShut {NoStop}%
\bibitem [{\citenamefont {Gustafsson}(2005)}]{Gustafsson2005}%
  \BibitemOpen
  \bibfield  {author} {\bibinfo {author} {\bibfnamefont {M.~G.~L.}\
  \bibnamefont {Gustafsson}},\ }\href
  {http://dx.doi.org/10.1073/pnas.0406877102} {\bibfield  {journal} {\bibinfo
  {journal} {Proceedings of the National Academy of Sciences}\ }\textbf
  {\bibinfo {volume} {102}},\ \bibinfo {pages} {13081} (\bibinfo {year}
  {2005})}\BibitemShut {NoStop}%
\bibitem [{\citenamefont {Chmyrov}\ \emph {et~al.}(2013)\citenamefont
  {Chmyrov}, \citenamefont {Keller}, \citenamefont {Grotjohann}, \citenamefont
  {Ratz}, \citenamefont {d'Este}, \citenamefont {Jakobs}, \citenamefont
  {Eggeling},\ and\ \citenamefont {Hell}}]{Chmyrov2013}%
  \BibitemOpen
  \bibfield  {author} {\bibinfo {author} {\bibfnamefont {A.}~\bibnamefont
  {Chmyrov}}, \bibinfo {author} {\bibfnamefont {J.}~\bibnamefont {Keller}},
  \bibinfo {author} {\bibfnamefont {T.}~\bibnamefont {Grotjohann}}, \bibinfo
  {author} {\bibfnamefont {M.}~\bibnamefont {Ratz}}, \bibinfo {author}
  {\bibfnamefont {E.}~\bibnamefont {d'Este}}, \bibinfo {author} {\bibfnamefont
  {S.}~\bibnamefont {Jakobs}}, \bibinfo {author} {\bibfnamefont
  {C.}~\bibnamefont {Eggeling}}, \ and\ \bibinfo {author} {\bibfnamefont
  {S.~W.}\ \bibnamefont {Hell}},\ }\href {http://dx.doi.org/10.1038/nmeth.2556}
  {\bibfield  {journal} {\bibinfo  {journal} {Nature Methods}\ }\textbf
  {\bibinfo {volume} {10}},\ \bibinfo {pages} {737} (\bibinfo {year}
  {2013})}\BibitemShut {NoStop}%
\bibitem [{\citenamefont {Thiel}\ \emph {et~al.}(2016)\citenamefont {Thiel},
  \citenamefont {Rohner}, \citenamefont {Ganzhorn}, \citenamefont {Appel},
  \citenamefont {Neu}, \citenamefont {M\"uller}, \citenamefont {Kleiner},
  \citenamefont {Koelle},\ and\ \citenamefont {Maletinsky}}]{Thiel2016}%
  \BibitemOpen
  \bibfield  {author} {\bibinfo {author} {\bibfnamefont {L.}~\bibnamefont
  {Thiel}}, \bibinfo {author} {\bibfnamefont {D.}~\bibnamefont {Rohner}},
  \bibinfo {author} {\bibfnamefont {M.}~\bibnamefont {Ganzhorn}}, \bibinfo
  {author} {\bibfnamefont {P.}~\bibnamefont {Appel}}, \bibinfo {author}
  {\bibfnamefont {E.}~\bibnamefont {Neu}}, \bibinfo {author} {\bibfnamefont
  {B.}~\bibnamefont {M\"uller}}, \bibinfo {author} {\bibfnamefont
  {R.}~\bibnamefont {Kleiner}}, \bibinfo {author} {\bibfnamefont
  {D.}~\bibnamefont {Koelle}}, \ and\ \bibinfo {author} {\bibfnamefont
  {P.}~\bibnamefont {Maletinsky}},\ }\href
  {http://dx.doi.org/10.1038/nnano.2016.63} {\bibfield  {journal} {\bibinfo
  {journal} {Nature Nanotechnology}\ }\textbf {\bibinfo {volume} {11}},\
  \bibinfo {pages} {677} (\bibinfo {year} {2016})}\BibitemShut {NoStop}%
\bibitem [{\citenamefont {Pelliccione}\ \emph {et~al.}(2016)\citenamefont
  {Pelliccione}, \citenamefont {Jenkins}, \citenamefont {Ovartchaiyapong},
  \citenamefont {Reetz}, \citenamefont {Emmanouilidou}, \citenamefont {Ni},\
  and\ \citenamefont {Jayich}}]{Pelliccione2016}%
  \BibitemOpen
  \bibfield  {author} {\bibinfo {author} {\bibfnamefont {M.}~\bibnamefont
  {Pelliccione}}, \bibinfo {author} {\bibfnamefont {A.}~\bibnamefont
  {Jenkins}}, \bibinfo {author} {\bibfnamefont {P.}~\bibnamefont
  {Ovartchaiyapong}}, \bibinfo {author} {\bibfnamefont {C.}~\bibnamefont
  {Reetz}}, \bibinfo {author} {\bibfnamefont {E.}~\bibnamefont
  {Emmanouilidou}}, \bibinfo {author} {\bibfnamefont {N.}~\bibnamefont {Ni}}, \
  and\ \bibinfo {author} {\bibfnamefont {A.~C.~B.}\ \bibnamefont {Jayich}},\
  }\href {http://dx.doi.org/10.1038/nnano.2016.68} {\bibfield  {journal}
  {\bibinfo  {journal} {Nature nanotechnology}\ }\textbf {\bibinfo {volume}
  {11}},\ \bibinfo {pages} {700} (\bibinfo {year} {2016})}\BibitemShut
  {NoStop}%
\bibitem [{\citenamefont {Grinolds}\ \emph {et~al.}(2014)\citenamefont
  {Grinolds}, \citenamefont {Warner}, \citenamefont {De~Greve}, \citenamefont
  {Dovzhenko}, \citenamefont {Thiel}, \citenamefont {{Walsworth}},
  \citenamefont {Hong}, \citenamefont {Maletinsky},\ and\ \citenamefont
  {Yacoby}}]{Grinolds2014}%
  \BibitemOpen
  \bibfield  {author} {\bibinfo {author} {\bibfnamefont {M.}~\bibnamefont
  {Grinolds}}, \bibinfo {author} {\bibfnamefont {M.}~\bibnamefont {Warner}},
  \bibinfo {author} {\bibfnamefont {K.}~\bibnamefont {De~Greve}}, \bibinfo
  {author} {\bibfnamefont {Y.}~\bibnamefont {Dovzhenko}}, \bibinfo {author}
  {\bibfnamefont {L.}~\bibnamefont {Thiel}}, \bibinfo {author} {\bibfnamefont
  {R.~L.~L.}\ \bibnamefont {{Walsworth}}}, \bibinfo {author} {\bibfnamefont
  {S.}~\bibnamefont {Hong}}, \bibinfo {author} {\bibfnamefont {P.}~\bibnamefont
  {Maletinsky}}, \ and\ \bibinfo {author} {\bibfnamefont {A.}~\bibnamefont
  {Yacoby}},\ }\href {http://dx.doi.org/10.1038/nnano.2014.30} {\bibfield
  {journal} {\bibinfo  {journal} {Nature Nanotechnology}\ }\textbf {\bibinfo
  {volume} {9}},\ \bibinfo {pages} {279} (\bibinfo {year} {2014})}\BibitemShut
  {NoStop}%
\bibitem [{\citenamefont {Appel}\ \emph
  {et~al.}(2016{\natexlab{a}})\citenamefont {Appel}, \citenamefont {Neu},
  \citenamefont {Ganzhorn}, \citenamefont {Barfuss}, \citenamefont {Batzer},
  \citenamefont {Gratz}, \citenamefont {Tsch{\"o}pe},\ and\ \citenamefont
  {Maletinsky}}]{fabricationMaletinsky}%
  \BibitemOpen
  \bibfield  {author} {\bibinfo {author} {\bibfnamefont {P.}~\bibnamefont
  {Appel}}, \bibinfo {author} {\bibfnamefont {E.}~\bibnamefont {Neu}}, \bibinfo
  {author} {\bibfnamefont {M.}~\bibnamefont {Ganzhorn}}, \bibinfo {author}
  {\bibfnamefont {A.}~\bibnamefont {Barfuss}}, \bibinfo {author} {\bibfnamefont
  {M.}~\bibnamefont {Batzer}}, \bibinfo {author} {\bibfnamefont
  {M.}~\bibnamefont {Gratz}}, \bibinfo {author} {\bibfnamefont
  {A.}~\bibnamefont {Tsch{\"o}pe}}, \ and\ \bibinfo {author} {\bibfnamefont
  {P.}~\bibnamefont {Maletinsky}},\ }\href@noop {} {\bibfield  {journal}
  {\bibinfo  {journal} {Review of Scientific Instruments}\ }\textbf {\bibinfo
  {volume} {87}},\ \bibinfo {pages} {063703} (\bibinfo {year}
  {2016}{\natexlab{a}})}\BibitemShut {NoStop}%
\bibitem [{\citenamefont {Bouchard}\ \emph {et~al.}(2011)\citenamefont
  {Bouchard}, \citenamefont {Acosta}, \citenamefont {Bauch},\ and\
  \citenamefont {Budker}}]{bouchard2011}%
  \BibitemOpen
  \bibfield  {author} {\bibinfo {author} {\bibfnamefont {L.-S.}\ \bibnamefont
  {Bouchard}}, \bibinfo {author} {\bibfnamefont {V.~M.}\ \bibnamefont
  {Acosta}}, \bibinfo {author} {\bibfnamefont {E.}~\bibnamefont {Bauch}}, \
  and\ \bibinfo {author} {\bibfnamefont {D.}~\bibnamefont {Budker}},\ }\href
  {https://doi.org/10.1088/1367-2630/13/2/025017} {\bibfield  {journal}
  {\bibinfo  {journal} {New Journal of Physics}\ }\textbf {\bibinfo {volume}
  {13}},\ \bibinfo {pages} {025017} (\bibinfo {year} {2011})}\BibitemShut
  {NoStop}%
\bibitem [{\citenamefont {Lim}\ and\ \citenamefont {Byrne}(1997)}]{Lim1997}%
  \BibitemOpen
  \bibfield  {author} {\bibinfo {author} {\bibfnamefont {H.~J.}\ \bibnamefont
  {Lim}}\ and\ \bibinfo {author} {\bibfnamefont {J.~G.}\ \bibnamefont
  {Byrne}},\ }\href {\doibase 10.1007/s11663-997-0108-1} {\bibfield  {journal}
  {\bibinfo  {journal} {Metallurgical and Materials Transactions B}\ }\textbf
  {\bibinfo {volume} {28}},\ \bibinfo {pages} {425} (\bibinfo {year}
  {1997})}\BibitemShut {NoStop}%
\bibitem [{\citenamefont {Acosta}\ \emph
  {et~al.}(2010{\natexlab{c}})\citenamefont {Acosta}, \citenamefont {Bauch},
  \citenamefont {Ledbetter}, \citenamefont {Waxman}, \citenamefont {Bouchard},\
  and\ \citenamefont {Budker}}]{Acosta2010}%
  \BibitemOpen
  \bibfield  {author} {\bibinfo {author} {\bibfnamefont {V.~M.}\ \bibnamefont
  {Acosta}}, \bibinfo {author} {\bibfnamefont {E.}~\bibnamefont {Bauch}},
  \bibinfo {author} {\bibfnamefont {M.~P.}\ \bibnamefont {Ledbetter}}, \bibinfo
  {author} {\bibfnamefont {A.}~\bibnamefont {Waxman}}, \bibinfo {author}
  {\bibfnamefont {L.-S.}\ \bibnamefont {Bouchard}}, \ and\ \bibinfo {author}
  {\bibfnamefont {D.}~\bibnamefont {Budker}},\ }\href
  {https://link.aps.org/doi/10.1103/PhysRevLett.104.070801} {\bibfield
  {journal} {\bibinfo  {journal} {Phys. Rev. Lett.}\ }\textbf {\bibinfo
  {volume} {104}},\ \bibinfo {pages} {070801} (\bibinfo {year}
  {2010}{\natexlab{c}})}\BibitemShut {NoStop}%
\bibitem [{\citenamefont {Chen}\ \emph {et~al.}(2011)\citenamefont {Chen},
  \citenamefont {Dong}, \citenamefont {Sun}, \citenamefont {Zou}, \citenamefont
  {Cui}, \citenamefont {Han},\ and\ \citenamefont {Guo}}]{Chen2011}%
  \BibitemOpen
  \bibfield  {author} {\bibinfo {author} {\bibfnamefont {X.-D.}\ \bibnamefont
  {Chen}}, \bibinfo {author} {\bibfnamefont {C.-H.}\ \bibnamefont {Dong}},
  \bibinfo {author} {\bibfnamefont {F.-W.}\ \bibnamefont {Sun}}, \bibinfo
  {author} {\bibfnamefont {C.-L.}\ \bibnamefont {Zou}}, \bibinfo {author}
  {\bibfnamefont {J.-M.}\ \bibnamefont {Cui}}, \bibinfo {author} {\bibfnamefont
  {Z.-F.}\ \bibnamefont {Han}}, \ and\ \bibinfo {author} {\bibfnamefont
  {G.-C.}\ \bibnamefont {Guo}},\ }\href {https://doi.org/10.1063/1.3652910}
  {\bibfield  {journal} {\bibinfo  {journal} {Applied Physics Letters}\
  }\textbf {\bibinfo {volume} {99}},\ \bibinfo {pages} {161903} (\bibinfo
  {year} {2011})}\BibitemShut {NoStop}%
\bibitem [{\citenamefont {Doherty}\ \emph
  {et~al.}(2014{\natexlab{a}})\citenamefont {Doherty}, \citenamefont {Acosta},
  \citenamefont {Jarmola}, \citenamefont {Barson}, \citenamefont {Manson},
  \citenamefont {Budker},\ and\ \citenamefont {Hollenberg}}]{Doherty2014}%
  \BibitemOpen
  \bibfield  {author} {\bibinfo {author} {\bibfnamefont {M.~W.}\ \bibnamefont
  {Doherty}}, \bibinfo {author} {\bibfnamefont {V.~M.}\ \bibnamefont {Acosta}},
  \bibinfo {author} {\bibfnamefont {A.}~\bibnamefont {Jarmola}}, \bibinfo
  {author} {\bibfnamefont {M.~S.~J.}\ \bibnamefont {Barson}}, \bibinfo {author}
  {\bibfnamefont {N.~B.}\ \bibnamefont {Manson}}, \bibinfo {author}
  {\bibfnamefont {D.}~\bibnamefont {Budker}}, \ and\ \bibinfo {author}
  {\bibfnamefont {L.~C.~L.}\ \bibnamefont {Hollenberg}},\ }\href {\doibase
  10.1103/PhysRevB.90.041201} {\bibfield  {journal} {\bibinfo  {journal} {Phys.
  Rev. B}\ }\textbf {\bibinfo {volume} {90}},\ \bibinfo {pages} {041201}
  (\bibinfo {year} {2014}{\natexlab{a}})}\BibitemShut {NoStop}%
\bibitem [{\citenamefont {Laraoui}\ \emph {et~al.}(2015)\citenamefont
  {Laraoui}, \citenamefont {Aycock-Rizzo}, \citenamefont {Gao}, \citenamefont
  {Lu}, \citenamefont {Riedo},\ and\ \citenamefont {Meriles}}]{Laraoui2015}%
  \BibitemOpen
  \bibfield  {author} {\bibinfo {author} {\bibfnamefont {A.}~\bibnamefont
  {Laraoui}}, \bibinfo {author} {\bibfnamefont {H.}~\bibnamefont
  {Aycock-Rizzo}}, \bibinfo {author} {\bibfnamefont {Y.}~\bibnamefont {Gao}},
  \bibinfo {author} {\bibfnamefont {X.}~\bibnamefont {Lu}}, \bibinfo {author}
  {\bibfnamefont {E.}~\bibnamefont {Riedo}}, \ and\ \bibinfo {author}
  {\bibfnamefont {C.~A.}\ \bibnamefont {Meriles}},\ }\href
  {http://dx.doi.org/10.1038/ncomms9954} {\bibfield  {journal} {\bibinfo
  {journal} {Nature Communications}\ }\textbf {\bibinfo {volume} {6}},\
  \bibinfo {pages} {8954 EP } (\bibinfo {year} {2015})}\BibitemShut {NoStop}%
\bibitem [{\citenamefont {Kucsko}\ \emph {et~al.}(2013)\citenamefont {Kucsko},
  \citenamefont {Maurer}, \citenamefont {Yao}, \citenamefont {Kubo},
  \citenamefont {Noh}, \citenamefont {Lo}, \citenamefont {Park},\ and\
  \citenamefont {Lukin}}]{Kucsko2013}%
  \BibitemOpen
  \bibfield  {author} {\bibinfo {author} {\bibfnamefont {G.}~\bibnamefont
  {Kucsko}}, \bibinfo {author} {\bibfnamefont {P.~C.}\ \bibnamefont {Maurer}},
  \bibinfo {author} {\bibfnamefont {N.~Y.}\ \bibnamefont {Yao}}, \bibinfo
  {author} {\bibfnamefont {M.}~\bibnamefont {Kubo}}, \bibinfo {author}
  {\bibfnamefont {H.~J.}\ \bibnamefont {Noh}}, \bibinfo {author} {\bibfnamefont
  {P.~K.}\ \bibnamefont {Lo}}, \bibinfo {author} {\bibfnamefont
  {H.}~\bibnamefont {Park}}, \ and\ \bibinfo {author} {\bibfnamefont {M.~D.}\
  \bibnamefont {Lukin}},\ }\href {http://dx.doi.org/10.1038/nature12373}
  {\bibfield  {journal} {\bibinfo  {journal} {Nature}\ }\textbf {\bibinfo
  {volume} {500}},\ \bibinfo {pages} {54 EP } (\bibinfo {year}
  {2013})}\BibitemShut {NoStop}%
\bibitem [{\citenamefont {Kehayias}\ \emph {et~al.}(2014)\citenamefont
  {Kehayias}, \citenamefont {Mr\'ozek}, \citenamefont {Acosta}, \citenamefont
  {Jarmola}, \citenamefont {Rudnicki}, \citenamefont {Folman}, \citenamefont
  {Gawlik},\ and\ \citenamefont {Budker}}]{Kehayias2014}%
  \BibitemOpen
  \bibfield  {author} {\bibinfo {author} {\bibfnamefont {P.}~\bibnamefont
  {Kehayias}}, \bibinfo {author} {\bibfnamefont {M.}~\bibnamefont {Mr\'ozek}},
  \bibinfo {author} {\bibfnamefont {V.~M.}\ \bibnamefont {Acosta}}, \bibinfo
  {author} {\bibfnamefont {A.}~\bibnamefont {Jarmola}}, \bibinfo {author}
  {\bibfnamefont {D.~S.}\ \bibnamefont {Rudnicki}}, \bibinfo {author}
  {\bibfnamefont {R.}~\bibnamefont {Folman}}, \bibinfo {author} {\bibfnamefont
  {W.}~\bibnamefont {Gawlik}}, \ and\ \bibinfo {author} {\bibfnamefont
  {D.}~\bibnamefont {Budker}},\ }\href {\doibase 10.1103/PhysRevB.89.245202}
  {\bibfield  {journal} {\bibinfo  {journal} {Phys. Rev. B}\ }\textbf {\bibinfo
  {volume} {89}},\ \bibinfo {pages} {245202} (\bibinfo {year}
  {2014})}\BibitemShut {NoStop}%
\bibitem [{\citenamefont {Fang}\ \emph {et~al.}(2013)\citenamefont {Fang},
  \citenamefont {Acosta}, \citenamefont {Santori}, \citenamefont {Huang},
  \citenamefont {Itoh}, \citenamefont {Watanabe}, \citenamefont {Shikata},\
  and\ \citenamefont {Beausoleil}}]{Fang2013}%
  \BibitemOpen
  \bibfield  {author} {\bibinfo {author} {\bibfnamefont {K.}~\bibnamefont
  {Fang}}, \bibinfo {author} {\bibfnamefont {V.~M.}\ \bibnamefont {Acosta}},
  \bibinfo {author} {\bibfnamefont {C.}~\bibnamefont {Santori}}, \bibinfo
  {author} {\bibfnamefont {Z.}~\bibnamefont {Huang}}, \bibinfo {author}
  {\bibfnamefont {K.~M.}\ \bibnamefont {Itoh}}, \bibinfo {author}
  {\bibfnamefont {H.}~\bibnamefont {Watanabe}}, \bibinfo {author}
  {\bibfnamefont {S.}~\bibnamefont {Shikata}}, \ and\ \bibinfo {author}
  {\bibfnamefont {R.~G.}\ \bibnamefont {Beausoleil}},\ }\href {\doibase
  10.1103/PhysRevLett.110.130802} {\bibfield  {journal} {\bibinfo  {journal}
  {Phys. Rev. Lett.}\ }\textbf {\bibinfo {volume} {110}},\ \bibinfo {pages}
  {130802} (\bibinfo {year} {2013})}\BibitemShut {NoStop}%
\bibitem [{\citenamefont {Waxman}\ \emph {et~al.}(2014)\citenamefont {Waxman},
  \citenamefont {Schlussel}, \citenamefont {Groswasser}, \citenamefont
  {Acosta}, \citenamefont {Bouchard}, \citenamefont {Budker},\ and\
  \citenamefont {Folman}}]{PRBwaxman}%
  \BibitemOpen
  \bibfield  {author} {\bibinfo {author} {\bibfnamefont {A.}~\bibnamefont
  {Waxman}}, \bibinfo {author} {\bibfnamefont {Y.}~\bibnamefont {Schlussel}},
  \bibinfo {author} {\bibfnamefont {D.}~\bibnamefont {Groswasser}}, \bibinfo
  {author} {\bibfnamefont {V.~M.}\ \bibnamefont {Acosta}}, \bibinfo {author}
  {\bibfnamefont {L.-S.}\ \bibnamefont {Bouchard}}, \bibinfo {author}
  {\bibfnamefont {D.}~\bibnamefont {Budker}}, \ and\ \bibinfo {author}
  {\bibfnamefont {R.}~\bibnamefont {Folman}},\ }\href {\doibase
  10.1103/PhysRevB.89.054509} {\bibfield  {journal} {\bibinfo  {journal}
  {Physical Review B}\ }\textbf {\bibinfo {volume} {89}},\ \bibinfo {pages}
  {054509} (\bibinfo {year} {2014})}\BibitemShut {NoStop}%
\bibitem [{\citenamefont {Alfasi}\ \emph {et~al.}(2016)\citenamefont {Alfasi},
  \citenamefont {Masis}, \citenamefont {Shtempluck}, \citenamefont {Kochetok},\
  and\ \citenamefont {Buks}}]{Alfasi2016}%
  \BibitemOpen
  \bibfield  {author} {\bibinfo {author} {\bibfnamefont {N.}~\bibnamefont
  {Alfasi}}, \bibinfo {author} {\bibfnamefont {S.}~\bibnamefont {Masis}},
  \bibinfo {author} {\bibfnamefont {O.}~\bibnamefont {Shtempluck}}, \bibinfo
  {author} {\bibfnamefont {V.}~\bibnamefont {Kochetok}}, \ and\ \bibinfo
  {author} {\bibfnamefont {E.}~\bibnamefont {Buks}},\ }\href
  {https://doi.org/10.1063/1.4959225} {\bibfield  {journal} {\bibinfo
  {journal} {AIP Advances}\ }\textbf {\bibinfo {volume} {6}},\ \bibinfo {pages}
  {075311} (\bibinfo {year} {2016})}\BibitemShut {NoStop}%
\bibitem [{\citenamefont {Schlussel}\ \emph {et~al.}(2018)\citenamefont
  {Schlussel}, \citenamefont {Lenz}, \citenamefont {Rohner}, \citenamefont
  {Bar-Haim}, \citenamefont {Bougas}, \citenamefont {Groswasser}, \citenamefont
  {Kieschnick}, \citenamefont {Rozenberg}, \citenamefont {Thiel}, \citenamefont
  {Waxman} \emph {et~al.}}]{schlussel2018widefield}%
  \BibitemOpen
  \bibfield  {author} {\bibinfo {author} {\bibfnamefont {Y.}~\bibnamefont
  {Schlussel}}, \bibinfo {author} {\bibfnamefont {T.}~\bibnamefont {Lenz}},
  \bibinfo {author} {\bibfnamefont {D.}~\bibnamefont {Rohner}}, \bibinfo
  {author} {\bibfnamefont {Y.}~\bibnamefont {Bar-Haim}}, \bibinfo {author}
  {\bibfnamefont {L.}~\bibnamefont {Bougas}}, \bibinfo {author} {\bibfnamefont
  {D.}~\bibnamefont {Groswasser}}, \bibinfo {author} {\bibfnamefont
  {M.}~\bibnamefont {Kieschnick}}, \bibinfo {author} {\bibfnamefont
  {E.}~\bibnamefont {Rozenberg}}, \bibinfo {author} {\bibfnamefont
  {L.}~\bibnamefont {Thiel}}, \bibinfo {author} {\bibfnamefont
  {A.}~\bibnamefont {Waxman}},  \emph {et~al.},\ }\href
  {https://arxiv.org/abs/1803.01957} {\bibfield  {journal} {\bibinfo  {journal}
  {arXiv preprint arXiv:1803.01957}\ } (\bibinfo {year} {2018})}\BibitemShut
  {NoStop}%
\bibitem [{\citenamefont {Nusran}\ \emph {et~al.}(2018)\citenamefont {Nusran},
  \citenamefont {Joshi}, \citenamefont {Cho}, \citenamefont {Tanatar},
  \citenamefont {Meier}, \citenamefont {Bud'ko}, \citenamefont {Canfield},
  \citenamefont {Liu}, \citenamefont {Lograsso},\ and\ \citenamefont
  {Prozorov}}]{Nusran2018}%
  \BibitemOpen
  \bibfield  {author} {\bibinfo {author} {\bibfnamefont {N.~M.}\ \bibnamefont
  {Nusran}}, \bibinfo {author} {\bibfnamefont {K.~R.}\ \bibnamefont {Joshi}},
  \bibinfo {author} {\bibfnamefont {K.}~\bibnamefont {Cho}}, \bibinfo {author}
  {\bibfnamefont {M.~A.}\ \bibnamefont {Tanatar}}, \bibinfo {author}
  {\bibfnamefont {W.~R.}\ \bibnamefont {Meier}}, \bibinfo {author}
  {\bibfnamefont {S.~L.}\ \bibnamefont {Bud'ko}}, \bibinfo {author}
  {\bibfnamefont {P.~C.}\ \bibnamefont {Canfield}}, \bibinfo {author}
  {\bibfnamefont {Y.}~\bibnamefont {Liu}}, \bibinfo {author} {\bibfnamefont
  {T.~A.}\ \bibnamefont {Lograsso}}, \ and\ \bibinfo {author} {\bibfnamefont
  {R.}~\bibnamefont {Prozorov}},\ }\href
  {http://stacks.iop.org/1367-2630/20/i=4/a=043010} {\bibfield  {journal}
  {\bibinfo  {journal} {New Journal of Physics}\ }\textbf {\bibinfo {volume}
  {20}},\ \bibinfo {pages} {043010} (\bibinfo {year} {2018})}\BibitemShut
  {NoStop}%
\bibitem [{\citenamefont {{Joshi}}\ \emph {et~al.}(2018)\citenamefont
  {{Joshi}}, \citenamefont {{Nusran}}, \citenamefont {{Cho}}, \citenamefont
  {{Tanatar}}, \citenamefont {{Meier}}, \citenamefont {{Bud'ko}}, \citenamefont
  {{Canfield}},\ and\ \citenamefont {{Prozorov}}}]{Joshi2018}%
  \BibitemOpen
  \bibfield  {author} {\bibinfo {author} {\bibfnamefont {K.~R.}\ \bibnamefont
  {{Joshi}}}, \bibinfo {author} {\bibfnamefont {N.~M.}\ \bibnamefont
  {{Nusran}}}, \bibinfo {author} {\bibfnamefont {K.}~\bibnamefont {{Cho}}},
  \bibinfo {author} {\bibfnamefont {M.~A.}\ \bibnamefont {{Tanatar}}}, \bibinfo
  {author} {\bibfnamefont {W.~R.}\ \bibnamefont {{Meier}}}, \bibinfo {author}
  {\bibfnamefont {S.~L.}\ \bibnamefont {{Bud'ko}}}, \bibinfo {author}
  {\bibfnamefont {P.~C.}\ \bibnamefont {{Canfield}}}, \ and\ \bibinfo {author}
  {\bibfnamefont {R.}~\bibnamefont {{Prozorov}}},\ }\href@noop {} {\bibfield
  {journal} {\bibinfo  {journal} {ArXiv e-prints}\ } (\bibinfo {year}
  {2018})},\ \Eprint {http://arxiv.org/abs/1806.10689} {arXiv:1806.10689
  [cond-mat.supr-con]} \BibitemShut {NoStop}%
\bibitem [{\citenamefont {Dr\'eau}\ \emph {et~al.}(2011)\citenamefont
  {Dr\'eau}, \citenamefont {Lesik}, \citenamefont {Rondin}, \citenamefont
  {Spinicelli}, \citenamefont {Arcizet}, \citenamefont {Roch},\ and\
  \citenamefont {Jacques}}]{Dreau2011}%
  \BibitemOpen
  \bibfield  {author} {\bibinfo {author} {\bibfnamefont {A.}~\bibnamefont
  {Dr\'eau}}, \bibinfo {author} {\bibfnamefont {M.}~\bibnamefont {Lesik}},
  \bibinfo {author} {\bibfnamefont {L.}~\bibnamefont {Rondin}}, \bibinfo
  {author} {\bibfnamefont {P.}~\bibnamefont {Spinicelli}}, \bibinfo {author}
  {\bibfnamefont {O.}~\bibnamefont {Arcizet}}, \bibinfo {author} {\bibfnamefont
  {J.-F.}\ \bibnamefont {Roch}}, \ and\ \bibinfo {author} {\bibfnamefont
  {V.}~\bibnamefont {Jacques}},\ }\href {\doibase 10.1103/PhysRevB.84.195204}
  {\bibfield  {journal} {\bibinfo  {journal} {Phys. Rev. B}\ }\textbf {\bibinfo
  {volume} {84}},\ \bibinfo {pages} {195204} (\bibinfo {year}
  {2011})}\BibitemShut {NoStop}%
\bibitem [{\citenamefont {Casola}\ \emph {et~al.}(2018)\citenamefont {Casola},
  \citenamefont {van~der Sar},\ and\ \citenamefont {Yacoby}}]{Casola2018}%
  \BibitemOpen
  \bibfield  {author} {\bibinfo {author} {\bibfnamefont {F.}~\bibnamefont
  {Casola}}, \bibinfo {author} {\bibfnamefont {T.}~\bibnamefont {van~der Sar}},
  \ and\ \bibinfo {author} {\bibfnamefont {A.}~\bibnamefont {Yacoby}},\ }\href
  {http://dx.doi.org/10.1038/natrevmats.2017.88} {\bibfield  {journal}
  {\bibinfo  {journal} {Nature Reviews Materials}\ }\textbf {\bibinfo {volume}
  {3}},\ \bibinfo {pages} {17088 EP } (\bibinfo {year} {2018})}\BibitemShut
  {NoStop}%
\bibitem [{\citenamefont {Lesik}\ \emph {et~al.}(2016)\citenamefont {Lesik},
  \citenamefont {Raatz}, \citenamefont {Tallaire}, \citenamefont {Spinicelli},
  \citenamefont {John}, \citenamefont {Achard}, \citenamefont {Gicquel},
  \citenamefont {Jacques}, \citenamefont {Roch}, \citenamefont {Meijer},\ and\
  \citenamefont {Pezzagna}}]{Lesik2016}%
  \BibitemOpen
  \bibfield  {author} {\bibinfo {author} {\bibfnamefont {M.}~\bibnamefont
  {Lesik}}, \bibinfo {author} {\bibfnamefont {N.}~\bibnamefont {Raatz}},
  \bibinfo {author} {\bibfnamefont {A.}~\bibnamefont {Tallaire}}, \bibinfo
  {author} {\bibfnamefont {P.}~\bibnamefont {Spinicelli}}, \bibinfo {author}
  {\bibfnamefont {R.}~\bibnamefont {John}}, \bibinfo {author} {\bibfnamefont
  {J.}~\bibnamefont {Achard}}, \bibinfo {author} {\bibfnamefont
  {A.}~\bibnamefont {Gicquel}}, \bibinfo {author} {\bibfnamefont
  {V.}~\bibnamefont {Jacques}}, \bibinfo {author} {\bibfnamefont
  {J.}~\bibnamefont {Roch}}, \bibinfo {author} {\bibfnamefont {J.}~\bibnamefont
  {Meijer}}, \ and\ \bibinfo {author} {\bibfnamefont {S.}~\bibnamefont
  {Pezzagna}},\ }\href {https://dx.doi.org/10.1002/pssa.201600219} {\bibfield
  {journal} {\bibinfo  {journal} {Physica Status Solidi (A)}\ }\textbf
  {\bibinfo {volume} {213}},\ \bibinfo {pages} {2594} (\bibinfo {year}
  {2016})}\BibitemShut {NoStop}%
\bibitem [{\citenamefont {Lesik}\ \emph {et~al.}(2014)\citenamefont {Lesik},
  \citenamefont {Tetienne}, \citenamefont {Tallaire}, \citenamefont {Achard},
  \citenamefont {Mille}, \citenamefont {Gicquel}, \citenamefont {Roch},\ and\
  \citenamefont {Jacques}}]{Lesik2014}%
  \BibitemOpen
  \bibfield  {author} {\bibinfo {author} {\bibfnamefont {M.}~\bibnamefont
  {Lesik}}, \bibinfo {author} {\bibfnamefont {J.-P.}\ \bibnamefont {Tetienne}},
  \bibinfo {author} {\bibfnamefont {A.}~\bibnamefont {Tallaire}}, \bibinfo
  {author} {\bibfnamefont {J.}~\bibnamefont {Achard}}, \bibinfo {author}
  {\bibfnamefont {V.}~\bibnamefont {Mille}}, \bibinfo {author} {\bibfnamefont
  {A.}~\bibnamefont {Gicquel}}, \bibinfo {author} {\bibfnamefont {J.-F.}\
  \bibnamefont {Roch}}, \ and\ \bibinfo {author} {\bibfnamefont
  {V.}~\bibnamefont {Jacques}},\ }\href {https://doi.org/10.1063/1.4869103}
  {\bibfield  {journal} {\bibinfo  {journal} {Applied Physics Letters}\
  }\textbf {\bibinfo {volume} {104}},\ \bibinfo {pages} {113107} (\bibinfo
  {year} {2014})}\BibitemShut {NoStop}%
\bibitem [{\citenamefont {Lesik}\ \emph {et~al.}(2015)\citenamefont {Lesik},
  \citenamefont {Plays}, \citenamefont {Tallaire}, \citenamefont {Achard},
  \citenamefont {Brinza}, \citenamefont {William}, \citenamefont {Chipaux},
  \citenamefont {Toraille}, \citenamefont {Debuisschert}, \citenamefont
  {Gicquel}, \citenamefont {Roch},\ and\ \citenamefont {Jacques}}]{Lesik2015}%
  \BibitemOpen
  \bibfield  {author} {\bibinfo {author} {\bibfnamefont {M.}~\bibnamefont
  {Lesik}}, \bibinfo {author} {\bibfnamefont {T.}~\bibnamefont {Plays}},
  \bibinfo {author} {\bibfnamefont {A.}~\bibnamefont {Tallaire}}, \bibinfo
  {author} {\bibfnamefont {J.}~\bibnamefont {Achard}}, \bibinfo {author}
  {\bibfnamefont {O.}~\bibnamefont {Brinza}}, \bibinfo {author} {\bibfnamefont
  {L.}~\bibnamefont {William}}, \bibinfo {author} {\bibfnamefont
  {M.}~\bibnamefont {Chipaux}}, \bibinfo {author} {\bibfnamefont
  {L.}~\bibnamefont {Toraille}}, \bibinfo {author} {\bibfnamefont
  {T.}~\bibnamefont {Debuisschert}}, \bibinfo {author} {\bibfnamefont
  {A.}~\bibnamefont {Gicquel}}, \bibinfo {author} {\bibfnamefont
  {J.}~\bibnamefont {Roch}}, \ and\ \bibinfo {author} {\bibfnamefont
  {V.}~\bibnamefont {Jacques}},\ }\href
  {https://doi.org/10.1016/j.diamond.2015.05.003} {\bibfield  {journal}
  {\bibinfo  {journal} {Diamond and Related Materials}\ }\textbf {\bibinfo
  {volume} {56}} (\bibinfo {year} {2015})}\BibitemShut {NoStop}%
\bibitem [{\citenamefont {Kleinsasser}\ \emph {et~al.}(2016)\citenamefont
  {Kleinsasser}, \citenamefont {Stanfield}, \citenamefont {Banks},
  \citenamefont {Zhu}, \citenamefont {Li}, \citenamefont {Acosta},
  \citenamefont {Watanabe}, \citenamefont {Itoh},\ and\ \citenamefont
  {Fu}}]{Kleinsasser2016}%
  \BibitemOpen
  \bibfield  {author} {\bibinfo {author} {\bibfnamefont {E.~E.}\ \bibnamefont
  {Kleinsasser}}, \bibinfo {author} {\bibfnamefont {M.~M.}\ \bibnamefont
  {Stanfield}}, \bibinfo {author} {\bibfnamefont {J.~K.~Q.}\ \bibnamefont
  {Banks}}, \bibinfo {author} {\bibfnamefont {Z.}~\bibnamefont {Zhu}}, \bibinfo
  {author} {\bibfnamefont {W.-D.}\ \bibnamefont {Li}}, \bibinfo {author}
  {\bibfnamefont {V.~M.}\ \bibnamefont {Acosta}}, \bibinfo {author}
  {\bibfnamefont {H.}~\bibnamefont {Watanabe}}, \bibinfo {author}
  {\bibfnamefont {K.~M.}\ \bibnamefont {Itoh}}, \ and\ \bibinfo {author}
  {\bibfnamefont {K.-M.~C.}\ \bibnamefont {Fu}},\ }\href
  {https://doi.org/10.1063/1.4949357} {\bibfield  {journal} {\bibinfo
  {journal} {Applied Physics Letters}\ }\textbf {\bibinfo {volume} {108}},\
  \bibinfo {pages} {202401} (\bibinfo {year} {2016})}\BibitemShut {NoStop}%
\bibitem [{\citenamefont {Rittweger}\ \emph {et~al.}(2009)\citenamefont
  {Rittweger}, \citenamefont {Han}, \citenamefont {Irvine}, \citenamefont
  {Eggeling},\ and\ \citenamefont {Hell}}]{Rittweger2009}%
  \BibitemOpen
  \bibfield  {author} {\bibinfo {author} {\bibfnamefont {E.}~\bibnamefont
  {Rittweger}}, \bibinfo {author} {\bibfnamefont {K.~Y.}\ \bibnamefont {Han}},
  \bibinfo {author} {\bibfnamefont {S.~E.}\ \bibnamefont {Irvine}}, \bibinfo
  {author} {\bibfnamefont {C.}~\bibnamefont {Eggeling}}, \ and\ \bibinfo
  {author} {\bibfnamefont {S.~W.}\ \bibnamefont {Hell}},\ }\href
  {http://dx.doi.org/10.1038/nphoton.2009.2} {\bibfield  {journal} {\bibinfo
  {journal} {Nature Photonics}\ }\textbf {\bibinfo {volume} {3}},\ \bibinfo
  {pages} {144} (\bibinfo {year} {2009})}\BibitemShut {NoStop}%
\bibitem [{\citenamefont {Tetienne}\ \emph {et~al.}(2014)\citenamefont
  {Tetienne}, \citenamefont {Hingant}, \citenamefont {Kim}, \citenamefont
  {Diez}, \citenamefont {Adam}, \citenamefont {Garcia}, \citenamefont {Roch},
  \citenamefont {Rohart}, \citenamefont {Thiaville}, \citenamefont
  {Ravelosona},\ and\ \citenamefont {Jacques}}]{Tetienne2014a}%
  \BibitemOpen
  \bibfield  {author} {\bibinfo {author} {\bibfnamefont {J.-P.}\ \bibnamefont
  {Tetienne}}, \bibinfo {author} {\bibfnamefont {T.}~\bibnamefont {Hingant}},
  \bibinfo {author} {\bibfnamefont {J.-V.}\ \bibnamefont {Kim}}, \bibinfo
  {author} {\bibfnamefont {L.~H.}\ \bibnamefont {Diez}}, \bibinfo {author}
  {\bibfnamefont {J.-P.}\ \bibnamefont {Adam}}, \bibinfo {author}
  {\bibfnamefont {K.}~\bibnamefont {Garcia}}, \bibinfo {author} {\bibfnamefont
  {J.-F.}\ \bibnamefont {Roch}}, \bibinfo {author} {\bibfnamefont
  {S.}~\bibnamefont {Rohart}}, \bibinfo {author} {\bibfnamefont
  {A.}~\bibnamefont {Thiaville}}, \bibinfo {author} {\bibfnamefont
  {D.}~\bibnamefont {Ravelosona}}, \ and\ \bibinfo {author} {\bibfnamefont
  {V.}~\bibnamefont {Jacques}},\ }\href
  {http://dx.doi.org/10.1126/science.1250113} {\bibfield  {journal} {\bibinfo
  {journal} {Science}\ }\textbf {\bibinfo {volume} {344}},\ \bibinfo {pages}
  {1366} (\bibinfo {year} {2014})}\BibitemShut {NoStop}%
\bibitem [{\citenamefont {Gross}\ \emph {et~al.}(2017)\citenamefont {Gross},
  \citenamefont {Akhtar}, \citenamefont {Garcia}, \citenamefont
  {Mart{\'\i}nez}, \citenamefont {Chouaieb}, \citenamefont {Garcia},
  \citenamefont {Carr{\'e}t{\'e}ro}, \citenamefont {Barth{\'e}l{\'e}my},
  \citenamefont {Appel}, \citenamefont {Maletinsky}, \citenamefont {Kim},
  \citenamefont {Chauleau}, \citenamefont {Jaouen}, \citenamefont {Viret},
  \citenamefont {Bibes}, \citenamefont {Fusil},\ and\ \citenamefont
  {Jacques}}]{Gross2017}%
  \BibitemOpen
  \bibfield  {author} {\bibinfo {author} {\bibfnamefont {I.}~\bibnamefont
  {Gross}}, \bibinfo {author} {\bibfnamefont {W.}~\bibnamefont {Akhtar}},
  \bibinfo {author} {\bibfnamefont {V.}~\bibnamefont {Garcia}}, \bibinfo
  {author} {\bibfnamefont {L.~J.}\ \bibnamefont {Mart{\'\i}nez}}, \bibinfo
  {author} {\bibfnamefont {S.}~\bibnamefont {Chouaieb}}, \bibinfo {author}
  {\bibfnamefont {K.}~\bibnamefont {Garcia}}, \bibinfo {author} {\bibfnamefont
  {C.}~\bibnamefont {Carr{\'e}t{\'e}ro}}, \bibinfo {author} {\bibfnamefont
  {A.}~\bibnamefont {Barth{\'e}l{\'e}my}}, \bibinfo {author} {\bibfnamefont
  {P.}~\bibnamefont {Appel}}, \bibinfo {author} {\bibfnamefont
  {P.}~\bibnamefont {Maletinsky}}, \bibinfo {author} {\bibfnamefont {J.~V.}\
  \bibnamefont {Kim}}, \bibinfo {author} {\bibfnamefont {J.~Y.}\ \bibnamefont
  {Chauleau}}, \bibinfo {author} {\bibfnamefont {N.}~\bibnamefont {Jaouen}},
  \bibinfo {author} {\bibfnamefont {M.}~\bibnamefont {Viret}}, \bibinfo
  {author} {\bibfnamefont {M.}~\bibnamefont {Bibes}}, \bibinfo {author}
  {\bibfnamefont {S.}~\bibnamefont {Fusil}}, \ and\ \bibinfo {author}
  {\bibfnamefont {V.}~\bibnamefont {Jacques}},\ }\href
  {http://dx.doi.org/10.1038/nature23656} {\bibfield  {journal} {\bibinfo
  {journal} {Nature}\ }\textbf {\bibinfo {volume} {549}},\ \bibinfo {pages}
  {252} (\bibinfo {year} {2017})}\BibitemShut {NoStop}%
\bibitem [{\citenamefont {Balasubramanian}\ \emph {et~al.}(2008)\citenamefont
  {Balasubramanian}, \citenamefont {Chan}, \citenamefont {Kolesov},
  \citenamefont {Al-Hmoud}, \citenamefont {Tisler}, \citenamefont {Shin},
  \citenamefont {Kim}, \citenamefont {Wojcik}, \citenamefont {Hemmer},
  \citenamefont {Hanke}, \citenamefont {Leitenstorfer}, \citenamefont
  {Bratschitsch}, \citenamefont {Jelezko},\ and\ \citenamefont
  {Wrachtrup}}]{Balasubramanian2008}%
  \BibitemOpen
  \bibfield  {author} {\bibinfo {author} {\bibfnamefont {G.}~\bibnamefont
  {Balasubramanian}}, \bibinfo {author} {\bibfnamefont {I.}~\bibnamefont
  {Chan}}, \bibinfo {author} {\bibfnamefont {R.}~\bibnamefont {Kolesov}},
  \bibinfo {author} {\bibfnamefont {M.}~\bibnamefont {Al-Hmoud}}, \bibinfo
  {author} {\bibfnamefont {J.}~\bibnamefont {Tisler}}, \bibinfo {author}
  {\bibfnamefont {C.}~\bibnamefont {Shin}}, \bibinfo {author} {\bibfnamefont
  {C.}~\bibnamefont {Kim}}, \bibinfo {author} {\bibfnamefont {A.}~\bibnamefont
  {Wojcik}}, \bibinfo {author} {\bibfnamefont {A.}~\bibnamefont {Hemmer},
  \bibfnamefont {P.R. amd~Krueger}}, \bibinfo {author} {\bibfnamefont
  {T.}~\bibnamefont {Hanke}}, \bibinfo {author} {\bibfnamefont
  {A.}~\bibnamefont {Leitenstorfer}}, \bibinfo {author} {\bibfnamefont
  {R.}~\bibnamefont {Bratschitsch}}, \bibinfo {author} {\bibfnamefont
  {F.}~\bibnamefont {Jelezko}}, \ and\ \bibinfo {author} {\bibfnamefont
  {J.}~\bibnamefont {Wrachtrup}},\ }\href
  {http://dx.doi.org/10.1038/nature07278} {\bibfield  {journal} {\bibinfo
  {journal} {Nature}\ }\textbf {\bibinfo {volume} {455}},\ \bibinfo {pages}
  {648} (\bibinfo {year} {2008})}\BibitemShut {NoStop}%
\bibitem [{\citenamefont {Rondin}\ \emph {et~al.}(2012)\citenamefont {Rondin},
  \citenamefont {Tetienne}, \citenamefont {Spinicelli}, \citenamefont
  {Dal~Savio}, \citenamefont {Karrai}, \citenamefont {Dantelle}, \citenamefont
  {Thiaville}, \citenamefont {Rohart}, \citenamefont {Roch},\ and\
  \citenamefont {Jacques}}]{Rondin2011}%
  \BibitemOpen
  \bibfield  {author} {\bibinfo {author} {\bibfnamefont {L.}~\bibnamefont
  {Rondin}}, \bibinfo {author} {\bibfnamefont {J.-P.}\ \bibnamefont
  {Tetienne}}, \bibinfo {author} {\bibfnamefont {P.}~\bibnamefont
  {Spinicelli}}, \bibinfo {author} {\bibfnamefont {C.}~\bibnamefont
  {Dal~Savio}}, \bibinfo {author} {\bibfnamefont {K.}~\bibnamefont {Karrai}},
  \bibinfo {author} {\bibfnamefont {G.}~\bibnamefont {Dantelle}}, \bibinfo
  {author} {\bibfnamefont {A.}~\bibnamefont {Thiaville}}, \bibinfo {author}
  {\bibfnamefont {S.}~\bibnamefont {Rohart}}, \bibinfo {author} {\bibfnamefont
  {J.-F.}\ \bibnamefont {Roch}}, \ and\ \bibinfo {author} {\bibfnamefont
  {V.}~\bibnamefont {Jacques}},\ }\href {https://doi.org/10.1063/1.3703128}
  {\bibfield  {journal} {\bibinfo  {journal} {Appl. Phys. Lett.}\ ,\ \bibinfo
  {pages} {153118}} (\bibinfo {year} {2012})}\BibitemShut {NoStop}%
\bibitem [{\citenamefont {Maletinsky}\ \emph {et~al.}(2012)\citenamefont
  {Maletinsky}, \citenamefont {Hong}, \citenamefont {Grinolds}, \citenamefont
  {Hausmann}, \citenamefont {Lukin}, \citenamefont {Walsworth}, \citenamefont
  {Loncar},\ and\ \citenamefont {Yacoby}}]{Maletinsky2012}%
  \BibitemOpen
  \bibfield  {author} {\bibinfo {author} {\bibfnamefont {P.}~\bibnamefont
  {Maletinsky}}, \bibinfo {author} {\bibfnamefont {S.}~\bibnamefont {Hong}},
  \bibinfo {author} {\bibfnamefont {M.}~\bibnamefont {Grinolds}}, \bibinfo
  {author} {\bibfnamefont {B.}~\bibnamefont {Hausmann}}, \bibinfo {author}
  {\bibfnamefont {M.}~\bibnamefont {Lukin}}, \bibinfo {author} {\bibfnamefont
  {R.}~\bibnamefont {Walsworth}}, \bibinfo {author} {\bibfnamefont
  {M.}~\bibnamefont {Loncar}}, \ and\ \bibinfo {author} {\bibfnamefont
  {A.}~\bibnamefont {Yacoby}},\ }\href
  {http://dx.doi.org/10.1038/NNANO.2012.50} {\bibfield  {journal} {\bibinfo
  {journal} {Nat. Nanotechnol.}\ }\textbf {\bibinfo {volume} {7}},\ \bibinfo
  {pages} {320} (\bibinfo {year} {2012})}\BibitemShut {NoStop}%
\bibitem [{\citenamefont {Grinolds}\ \emph {et~al.}(2013)\citenamefont
  {Grinolds}, \citenamefont {Hong}, \citenamefont {Maletinsky}, \citenamefont
  {Luan}, \citenamefont {Lukin}, \citenamefont {Walsworth},\ and\ \citenamefont
  {Yacoby}}]{Grinolds2013}%
  \BibitemOpen
  \bibfield  {author} {\bibinfo {author} {\bibfnamefont {M.~S.}\ \bibnamefont
  {Grinolds}}, \bibinfo {author} {\bibfnamefont {S.}~\bibnamefont {Hong}},
  \bibinfo {author} {\bibfnamefont {P.}~\bibnamefont {Maletinsky}}, \bibinfo
  {author} {\bibfnamefont {L.}~\bibnamefont {Luan}}, \bibinfo {author}
  {\bibfnamefont {M.~D.}\ \bibnamefont {Lukin}}, \bibinfo {author}
  {\bibfnamefont {R.~L.}\ \bibnamefont {Walsworth}}, \ and\ \bibinfo {author}
  {\bibfnamefont {A.}~\bibnamefont {Yacoby}},\ }\href
  {http://dx.doi.org/10.1038/nphys2543} {\bibfield  {journal} {\bibinfo
  {journal} {Nature Physics}\ }\textbf {\bibinfo {volume} {9}},\ \bibinfo
  {pages} {215} (\bibinfo {year} {2013})}\BibitemShut {NoStop}%
\bibitem [{\citenamefont {Babinec}\ \emph {et~al.}(2010)\citenamefont
  {Babinec}, \citenamefont {Hausmann}, \citenamefont {Khan}, \citenamefont
  {Zhang}, \citenamefont {Maze}, \citenamefont {Hemmer},\ and\ \citenamefont
  {Lon\v{c}ar}}]{Babinec2010}%
  \BibitemOpen
  \bibfield  {author} {\bibinfo {author} {\bibfnamefont {T.~M.}\ \bibnamefont
  {Babinec}}, \bibinfo {author} {\bibfnamefont {B.~J.}\ \bibnamefont
  {Hausmann}}, \bibinfo {author} {\bibfnamefont {M.}~\bibnamefont {Khan}},
  \bibinfo {author} {\bibfnamefont {Y.}~\bibnamefont {Zhang}}, \bibinfo
  {author} {\bibfnamefont {J.~R.}\ \bibnamefont {Maze}}, \bibinfo {author}
  {\bibfnamefont {P.~R.}\ \bibnamefont {Hemmer}}, \ and\ \bibinfo {author}
  {\bibfnamefont {M.}~\bibnamefont {Lon\v{c}ar}},\ }\href
  {https://doi.org/10.1038/nnano.2010.6} {\bibfield  {journal} {\bibinfo
  {journal} {Nature Nanotechnology}\ }\textbf {\bibinfo {volume} {5}},\
  \bibinfo {pages} {195} (\bibinfo {year} {2010})}\BibitemShut {NoStop}%
\bibitem [{\citenamefont {Roth}\ \emph {et~al.}(1989)\citenamefont {Roth},
  \citenamefont {Sepulveda},\ and\ \citenamefont {Wikswo}}]{Roth1989}%
  \BibitemOpen
  \bibfield  {author} {\bibinfo {author} {\bibfnamefont {B.~J.}\ \bibnamefont
  {Roth}}, \bibinfo {author} {\bibfnamefont {N.~G.}\ \bibnamefont {Sepulveda}},
  \ and\ \bibinfo {author} {\bibfnamefont {J.~P.}\ \bibnamefont {Wikswo}},\
  }\href {http://dx.doi.org/10.1063/1.342549} {\bibfield  {journal} {\bibinfo
  {journal} {Journal of Applied Physics}\ }\textbf {\bibinfo {volume} {65}},\
  \bibinfo {pages} {361} (\bibinfo {year} {1989})}\BibitemShut {NoStop}%
\bibitem [{\citenamefont {Appel}\ \emph
  {et~al.}(2016{\natexlab{b}})\citenamefont {Appel}, \citenamefont {Neu},
  \citenamefont {Ganzhorn}, \citenamefont {Barfuss}, \citenamefont {Marietta},
  \citenamefont {Batzer}, \citenamefont {Gratz}, \citenamefont {Tsch\"ope},\
  and\ \citenamefont {Maletinsky}}]{Appel2016}%
  \BibitemOpen
  \bibfield  {author} {\bibinfo {author} {\bibfnamefont {P.}~\bibnamefont
  {Appel}}, \bibinfo {author} {\bibfnamefont {E.}~\bibnamefont {Neu}}, \bibinfo
  {author} {\bibfnamefont {M.}~\bibnamefont {Ganzhorn}}, \bibinfo {author}
  {\bibfnamefont {A.}~\bibnamefont {Barfuss}}, \bibinfo {author} {\bibnamefont
  {Marietta}}, \bibinfo {author} {\bibnamefont {Batzer}}, \bibinfo {author}
  {\bibfnamefont {M.}~\bibnamefont {Gratz}}, \bibinfo {author} {\bibfnamefont
  {A.}~\bibnamefont {Tsch\"ope}}, \ and\ \bibinfo {author} {\bibfnamefont
  {P.}~\bibnamefont {Maletinsky}},\ }\href {\doibase
  http://dx.doi.org/10.1063/1.4952953} {\bibfield  {journal} {\bibinfo
  {journal} {Review of Scientific Instruments}\ }\textbf {\bibinfo {volume}
  {87}},\ \bibinfo {pages} {063703} (\bibinfo {year}
  {2016}{\natexlab{b}})}\BibitemShut {NoStop}%
\bibitem [{\citenamefont {Blatter}\ \emph {et~al.}(1994)\citenamefont
  {Blatter}, \citenamefont {Feigel'man}, \citenamefont {Geshkenbein},
  \citenamefont {Larkin},\ and\ \citenamefont {Vinokur}}]{Blatter1994}%
  \BibitemOpen
  \bibfield  {author} {\bibinfo {author} {\bibfnamefont {G.}~\bibnamefont
  {Blatter}}, \bibinfo {author} {\bibfnamefont {M.~V.}\ \bibnamefont
  {Feigel'man}}, \bibinfo {author} {\bibfnamefont {V.~B.}\ \bibnamefont
  {Geshkenbein}}, \bibinfo {author} {\bibfnamefont {A.~I.}\ \bibnamefont
  {Larkin}}, \ and\ \bibinfo {author} {\bibfnamefont {V.~M.}\ \bibnamefont
  {Vinokur}},\ }\href {http://dx.doi.org/10.1103/RevModPhys.66.1125} {\bibfield
   {journal} {\bibinfo  {journal} {Rev. Mod. Phys.}\ }\textbf {\bibinfo
  {volume} {66}},\ \bibinfo {pages} {1125} (\bibinfo {year}
  {1994})}\BibitemShut {NoStop}%
\bibitem [{\citenamefont {Robledo}\ \emph
  {et~al.}(2011{\natexlab{b}})\citenamefont {Robledo}, \citenamefont
  {Childress}, \citenamefont {Bernien}, \citenamefont {Hensen}, \citenamefont
  {Alkemade},\ and\ \citenamefont {Hanson}}]{Robledo2011}%
  \BibitemOpen
  \bibfield  {author} {\bibinfo {author} {\bibfnamefont {L.}~\bibnamefont
  {Robledo}}, \bibinfo {author} {\bibfnamefont {L.}~\bibnamefont {Childress}},
  \bibinfo {author} {\bibfnamefont {H.}~\bibnamefont {Bernien}}, \bibinfo
  {author} {\bibfnamefont {B.}~\bibnamefont {Hensen}}, \bibinfo {author}
  {\bibfnamefont {P.~F.~A.}\ \bibnamefont {Alkemade}}, \ and\ \bibinfo {author}
  {\bibfnamefont {R.}~\bibnamefont {Hanson}},\ }\href
  {Http://dx.doi.org/10.1038/nature10401} {\bibfield  {journal} {\bibinfo
  {journal} {Nature}\ }\textbf {\bibinfo {volume} {477}},\ \bibinfo {pages}
  {574} (\bibinfo {year} {2011}{\natexlab{b}})}\BibitemShut {NoStop}%
\bibitem [{\citenamefont {Yale}\ \emph {et~al.}(2013)\citenamefont {Yale},
  \citenamefont {Buckley}, \citenamefont {Christle}, \citenamefont {Burkard},
  \citenamefont {Heremans}, \citenamefont {Bassett},\ and\ \citenamefont
  {Awschalom}}]{Yale2013}%
  \BibitemOpen
  \bibfield  {author} {\bibinfo {author} {\bibfnamefont {C.~G.}\ \bibnamefont
  {Yale}}, \bibinfo {author} {\bibfnamefont {B.~B.}\ \bibnamefont {Buckley}},
  \bibinfo {author} {\bibfnamefont {D.~J.}\ \bibnamefont {Christle}}, \bibinfo
  {author} {\bibfnamefont {G.}~\bibnamefont {Burkard}}, \bibinfo {author}
  {\bibfnamefont {F.~J.}\ \bibnamefont {Heremans}}, \bibinfo {author}
  {\bibfnamefont {L.~C.}\ \bibnamefont {Bassett}}, \ and\ \bibinfo {author}
  {\bibfnamefont {D.~D.}\ \bibnamefont {Awschalom}},\ }\href
  {http://dx.doi.org/10.1073/pnas.1305920110} {\bibfield  {journal} {\bibinfo
  {journal} {Proceedings of the National Academy of Sciences}\ }\textbf
  {\bibinfo {volume} {110}},\ \bibinfo {pages} {7595} (\bibinfo {year}
  {2013})}\BibitemShut {NoStop}%
\bibitem [{\citenamefont {Auslaender}\ \emph {et~al.}(2008)\citenamefont
  {Auslaender}, \citenamefont {Luan}, \citenamefont {Straver}, \citenamefont
  {Hoffman}, \citenamefont {Koshnick}, \citenamefont {Zeldov}, \citenamefont
  {Bonn}, \citenamefont {Liang}, \citenamefont {Hardy},\ and\ \citenamefont
  {Moler}}]{Auslaender2008}%
  \BibitemOpen
  \bibfield  {author} {\bibinfo {author} {\bibfnamefont {O.~M.}\ \bibnamefont
  {Auslaender}}, \bibinfo {author} {\bibfnamefont {L.}~\bibnamefont {Luan}},
  \bibinfo {author} {\bibfnamefont {E.~W.~J.}\ \bibnamefont {Straver}},
  \bibinfo {author} {\bibfnamefont {J.~E.}\ \bibnamefont {Hoffman}}, \bibinfo
  {author} {\bibfnamefont {N.~C.}\ \bibnamefont {Koshnick}}, \bibinfo {author}
  {\bibfnamefont {E.}~\bibnamefont {Zeldov}}, \bibinfo {author} {\bibfnamefont
  {D.~A.}\ \bibnamefont {Bonn}}, \bibinfo {author} {\bibfnamefont
  {R.}~\bibnamefont {Liang}}, \bibinfo {author} {\bibfnamefont {W.~N.}\
  \bibnamefont {Hardy}}, \ and\ \bibinfo {author} {\bibfnamefont {K.~A.}\
  \bibnamefont {Moler}},\ }\href {http://dx.doi.org/10.1038/nphys1127}
  {\bibfield  {journal} {\bibinfo  {journal} {Nature Physics}\ }\textbf
  {\bibinfo {volume} {5}},\ \bibinfo {pages} {35} (\bibinfo {year}
  {2008})}\BibitemShut {NoStop}%
\bibitem [{\citenamefont {Pearl}(1964)}]{Pearl1964}%
  \BibitemOpen
  \bibfield  {author} {\bibinfo {author} {\bibfnamefont {J.}~\bibnamefont
  {Pearl}},\ }\href {http://dx.doi.org/10.1063/1.1754056} {\bibfield  {journal}
  {\bibinfo  {journal} {Applied Physics Letters}\ }\textbf {\bibinfo {volume}
  {5}},\ \bibinfo {pages} {65} (\bibinfo {year} {1964})}\BibitemShut {NoStop}%
\bibitem [{\citenamefont {Carneiro}\ and\ \citenamefont
  {Brandt}(2000)}]{Carneiro2000}%
  \BibitemOpen
  \bibfield  {author} {\bibinfo {author} {\bibfnamefont {G.}~\bibnamefont
  {Carneiro}}\ and\ \bibinfo {author} {\bibfnamefont {E.~H.}\ \bibnamefont
  {Brandt}},\ }\href {http://dx.doi.org/10.1103/PhysRevB.61.6370} {\bibfield
  {journal} {\bibinfo  {journal} {Phys. Rev. B}\ }\textbf {\bibinfo {volume}
  {61}},\ \bibinfo {pages} {6370} (\bibinfo {year} {2000})}\BibitemShut
  {NoStop}%
\bibitem [{\citenamefont {Pezzagna}\ \emph {et~al.}(2010)\citenamefont
  {Pezzagna}, \citenamefont {Naydenoc}, \citenamefont {Jelezko}, \citenamefont
  {Wrachtrup},\ and\ \citenamefont {Meijer}}]{Pezzagna2010}%
  \BibitemOpen
  \bibfield  {author} {\bibinfo {author} {\bibfnamefont {S.}~\bibnamefont
  {Pezzagna}}, \bibinfo {author} {\bibfnamefont {B.}~\bibnamefont {Naydenoc}},
  \bibinfo {author} {\bibfnamefont {F.}~\bibnamefont {Jelezko}}, \bibinfo
  {author} {\bibfnamefont {J.}~\bibnamefont {Wrachtrup}}, \ and\ \bibinfo
  {author} {\bibfnamefont {J.}~\bibnamefont {Meijer}},\ }\href {\doibase
  0.1088/1367-2630/12/6/065017} {\bibfield  {journal} {\bibinfo  {journal} {New
  Journal of Physics}\ }\textbf {\bibinfo {volume} {12}},\ \bibinfo {pages}
  {065017} (\bibinfo {year} {2010})}\BibitemShut {NoStop}%
\bibitem [{\citenamefont {Myers}\ \emph {et~al.}(2014)\citenamefont {Myers},
  \citenamefont {Das}, \citenamefont {Dartiailh}, \citenamefont {Ohno},
  \citenamefont {Awschalom},\ and\ \citenamefont
  {Bleszynski~Jayich}}]{Myers2014}%
  \BibitemOpen
  \bibfield  {author} {\bibinfo {author} {\bibfnamefont {B.~A.}\ \bibnamefont
  {Myers}}, \bibinfo {author} {\bibfnamefont {A.}~\bibnamefont {Das}}, \bibinfo
  {author} {\bibfnamefont {M.~C.}\ \bibnamefont {Dartiailh}}, \bibinfo {author}
  {\bibfnamefont {K.}~\bibnamefont {Ohno}}, \bibinfo {author} {\bibfnamefont
  {D.~D.}\ \bibnamefont {Awschalom}}, \ and\ \bibinfo {author} {\bibfnamefont
  {A.~C.}\ \bibnamefont {Bleszynski~Jayich}},\ }\href
  {https://dx.doi.org/10.1103/PhysRevLett.113.027602} {\bibfield  {journal}
  {\bibinfo  {journal} {Phys. Rev. Lett.}\ }\textbf {\bibinfo {volume} {113}},\
  \bibinfo {pages} {027602} (\bibinfo {year} {2014})}\BibitemShut {NoStop}%
\bibitem [{\citenamefont {Shields}(2018)}]{Shields2018}%
  \BibitemOpen
  \bibfield  {author} {\bibinfo {author} {\bibfnamefont {B.}~\bibnamefont
  {Shields}},\ }\href@noop {} {\bibfield  {journal} {\bibinfo  {journal} {{\em
  private communication}}\ } (\bibinfo {year} {2018})}\BibitemShut {NoStop}%
\bibitem [{\citenamefont {Fischer}\ \emph {et~al.}(2007)\citenamefont
  {Fischer}, \citenamefont {Kugler}, \citenamefont {Maggio-Aprile},
  \citenamefont {Berthod},\ and\ \citenamefont {Renner}}]{Fischer2007}%
  \BibitemOpen
  \bibfield  {author} {\bibinfo {author} {\bibfnamefont {O.}~\bibnamefont
  {Fischer}}, \bibinfo {author} {\bibfnamefont {M.}~\bibnamefont {Kugler}},
  \bibinfo {author} {\bibfnamefont {I.}~\bibnamefont {Maggio-Aprile}}, \bibinfo
  {author} {\bibfnamefont {C.}~\bibnamefont {Berthod}}, \ and\ \bibinfo
  {author} {\bibfnamefont {C.}~\bibnamefont {Renner}},\ }\href
  {http://dx.doi.org/10.1103/RevModPhys.79.353} {\bibfield  {journal} {\bibinfo
   {journal} {Rev. Mod. Phys.}\ }\textbf {\bibinfo {volume} {79}},\ \bibinfo
  {pages} {353} (\bibinfo {year} {2007})}\BibitemShut {NoStop}%
\bibitem [{\citenamefont {Kirtley}(2010)}]{Kirtley2010}%
  \BibitemOpen
  \bibfield  {author} {\bibinfo {author} {\bibfnamefont {J.~R.}\ \bibnamefont
  {Kirtley}},\ }\href {\doibase 10.1088/0034-4885/73/12/126501} {\bibfield
  {journal} {\bibinfo  {journal} {Rep. Prog. Phys.}\ }\textbf {\bibinfo
  {volume} {73}},\ \bibinfo {pages} {126501} (\bibinfo {year}
  {2010})}\BibitemShut {NoStop}%
\bibitem [{\citenamefont {Suderow}\ \emph {et~al.}(2014)\citenamefont
  {Suderow}, \citenamefont {Guillam{\'o}n}, \citenamefont {Rodrigo},\ and\
  \citenamefont {Vieira}}]{Suderow2014}%
  \BibitemOpen
  \bibfield  {author} {\bibinfo {author} {\bibfnamefont {H.}~\bibnamefont
  {Suderow}}, \bibinfo {author} {\bibfnamefont {I.}~\bibnamefont
  {Guillam{\'o}n}}, \bibinfo {author} {\bibfnamefont {J.~G.}\ \bibnamefont
  {Rodrigo}}, \ and\ \bibinfo {author} {\bibfnamefont {S.}~\bibnamefont
  {Vieira}},\ }\href {http://stacks.iop.org/0953-2048/27/i=6/a=063001}
  {\bibfield  {journal} {\bibinfo  {journal} {Superconductor Science and
  Technology}\ }\textbf {\bibinfo {volume} {27}},\ \bibinfo {pages} {063001}
  (\bibinfo {year} {2014})}\BibitemShut {NoStop}%
\bibitem [{\citenamefont {Vasyukov}\ \emph {et~al.}(2013)\citenamefont
  {Vasyukov}, \citenamefont {Anahory}, \citenamefont {Embon}, \citenamefont
  {Halbertal}, \citenamefont {Cuppens}, \citenamefont {Neeman}, \citenamefont
  {Finkler}, \citenamefont {Segev}, \citenamefont {Myasoedov}, \citenamefont
  {Rappaport}, \citenamefont {Huber},\ and\ \citenamefont
  {Zeldov}}]{Vasyukov2013}%
  \BibitemOpen
  \bibfield  {author} {\bibinfo {author} {\bibfnamefont {D.}~\bibnamefont
  {Vasyukov}}, \bibinfo {author} {\bibfnamefont {Y.}~\bibnamefont {Anahory}},
  \bibinfo {author} {\bibfnamefont {L.}~\bibnamefont {Embon}}, \bibinfo
  {author} {\bibfnamefont {D.}~\bibnamefont {Halbertal}}, \bibinfo {author}
  {\bibfnamefont {J.}~\bibnamefont {Cuppens}}, \bibinfo {author} {\bibfnamefont
  {L.}~\bibnamefont {Neeman}}, \bibinfo {author} {\bibfnamefont
  {A.}~\bibnamefont {Finkler}}, \bibinfo {author} {\bibfnamefont
  {Y.}~\bibnamefont {Segev}}, \bibinfo {author} {\bibfnamefont
  {Y.}~\bibnamefont {Myasoedov}}, \bibinfo {author} {\bibfnamefont {M.~L.}\
  \bibnamefont {Rappaport}}, \bibinfo {author} {\bibfnamefont {M.~E.}\
  \bibnamefont {Huber}}, \ and\ \bibinfo {author} {\bibfnamefont
  {E.}~\bibnamefont {Zeldov}},\ }\href
  {http://dx.doi.org/10.1038/nnano.2013.169} {\bibfield  {journal} {\bibinfo
  {journal} {Nature Nanotechnology}\ }\textbf {\bibinfo {volume} {8}},\
  \bibinfo {pages} {639} (\bibinfo {year} {2013})}\BibitemShut {NoStop}%
\bibitem [{\citenamefont {Embon}\ \emph {et~al.}(2015)\citenamefont {Embon},
  \citenamefont {Anahory}, \citenamefont {Suhov}, \citenamefont {Halbertal},
  \citenamefont {Cuppens}, \citenamefont {Yakovenko}, \citenamefont {Uri},
  \citenamefont {Myasoedov}, \citenamefont {Rappaport}, \citenamefont {Huber}
  \emph {et~al.}}]{Embon2015}%
  \BibitemOpen
  \bibfield  {author} {\bibinfo {author} {\bibfnamefont {L.}~\bibnamefont
  {Embon}}, \bibinfo {author} {\bibfnamefont {Y.}~\bibnamefont {Anahory}},
  \bibinfo {author} {\bibfnamefont {A.}~\bibnamefont {Suhov}}, \bibinfo
  {author} {\bibfnamefont {D.}~\bibnamefont {Halbertal}}, \bibinfo {author}
  {\bibfnamefont {J.}~\bibnamefont {Cuppens}}, \bibinfo {author} {\bibfnamefont
  {A.}~\bibnamefont {Yakovenko}}, \bibinfo {author} {\bibfnamefont
  {A.}~\bibnamefont {Uri}}, \bibinfo {author} {\bibfnamefont {Y.}~\bibnamefont
  {Myasoedov}}, \bibinfo {author} {\bibfnamefont {M.~L.}\ \bibnamefont
  {Rappaport}}, \bibinfo {author} {\bibfnamefont {M.~E.}\ \bibnamefont
  {Huber}},  \emph {et~al.},\ }\href {http://dx.doi.org/10.1038/srep07598}
  {\bibfield  {journal} {\bibinfo  {journal} {Scientific reports}\ }\textbf
  {\bibinfo {volume} {5}},\ \bibinfo {pages} {7598} (\bibinfo {year}
  {2015})}\BibitemShut {NoStop}%
\bibitem [{\citenamefont {Embon}\ \emph
  {et~al.}(2017{\natexlab{a}})\citenamefont {Embon}, \citenamefont {Anahory},
  \citenamefont {Jeli{\'c}}, \citenamefont {Lachman}, \citenamefont
  {Myasoedov}, \citenamefont {Huber}, \citenamefont {Mikitik}, \citenamefont
  {Silhanek}, \citenamefont {Milo{\v{s}}evi{\'c}}, \citenamefont {Gurevich}
  \emph {et~al.}}]{Embon2017}%
  \BibitemOpen
  \bibfield  {author} {\bibinfo {author} {\bibfnamefont {L.}~\bibnamefont
  {Embon}}, \bibinfo {author} {\bibfnamefont {Y.}~\bibnamefont {Anahory}},
  \bibinfo {author} {\bibfnamefont {{\v{Z}}.~L.}\ \bibnamefont {Jeli{\'c}}},
  \bibinfo {author} {\bibfnamefont {E.~O.}\ \bibnamefont {Lachman}}, \bibinfo
  {author} {\bibfnamefont {Y.}~\bibnamefont {Myasoedov}}, \bibinfo {author}
  {\bibfnamefont {M.~E.}\ \bibnamefont {Huber}}, \bibinfo {author}
  {\bibfnamefont {G.~P.}\ \bibnamefont {Mikitik}}, \bibinfo {author}
  {\bibfnamefont {A.~V.}\ \bibnamefont {Silhanek}}, \bibinfo {author}
  {\bibfnamefont {M.~V.}\ \bibnamefont {Milo{\v{s}}evi{\'c}}}, \bibinfo
  {author} {\bibfnamefont {A.}~\bibnamefont {Gurevich}},  \emph {et~al.},\
  }\href {http://dx.doi.org/10.1038/s41467-017-00089-3} {\bibfield  {journal}
  {\bibinfo  {journal} {Nature communications}\ }\textbf {\bibinfo {volume}
  {8}},\ \bibinfo {pages} {85} (\bibinfo {year}
  {2017}{\natexlab{a}})}\BibitemShut {NoStop}%
\bibitem [{\citenamefont {Degen}\ \emph {et~al.}(2017)\citenamefont {Degen},
  \citenamefont {Reinhard},\ and\ \citenamefont {Cappellaro}}]{Degen2017}%
  \BibitemOpen
  \bibfield  {author} {\bibinfo {author} {\bibfnamefont {C.~L.}\ \bibnamefont
  {Degen}}, \bibinfo {author} {\bibfnamefont {F.}~\bibnamefont {Reinhard}}, \
  and\ \bibinfo {author} {\bibfnamefont {P.}~\bibnamefont {Cappellaro}},\
  }\href {\doibase 10.1103/RevModPhys.89.035002} {\bibfield  {journal}
  {\bibinfo  {journal} {Reviews of Modern Physics}\ }\textbf {\bibinfo {volume}
  {89}},\ \bibinfo {pages} {035002} (\bibinfo {year} {2017})}\BibitemShut
  {NoStop}%
\bibitem [{\citenamefont {Zeinali}\ \emph {et~al.}(2016)\citenamefont
  {Zeinali}, \citenamefont {Golod},\ and\ \citenamefont
  {Krasnov}}]{Zeinali2016}%
  \BibitemOpen
  \bibfield  {author} {\bibinfo {author} {\bibfnamefont {A.}~\bibnamefont
  {Zeinali}}, \bibinfo {author} {\bibfnamefont {T.}~\bibnamefont {Golod}}, \
  and\ \bibinfo {author} {\bibfnamefont {V.~M.}\ \bibnamefont {Krasnov}},\
  }\href {\doibase 10.1103/PhysRevB.94.214506} {\bibfield  {journal} {\bibinfo
  {journal} {Phys. Rev. B}\ }\textbf {\bibinfo {volume} {94}},\ \bibinfo
  {pages} {214506} (\bibinfo {year} {2016})}\BibitemShut {NoStop}%
\bibitem [{\citenamefont {Embon}\ \emph
  {et~al.}(2017{\natexlab{b}})\citenamefont {Embon}, \citenamefont {Anahory},
  \citenamefont {\u{Z}. L.~Jeli\'{c}}, \citenamefont {Lachman}, \citenamefont
  {Myasoedov}, \citenamefont {Huber}, \citenamefont {Mikitik}, \citenamefont
  {Silhanek}, \citenamefont {\u{s}evi\'{c}}, \citenamefont {aa. Gurevich},\
  and\ \citenamefont {Zeldov}}]{vortexdynamicszeldov2017}%
  \BibitemOpen
  \bibfield  {author} {\bibinfo {author} {\bibfnamefont {L.}~\bibnamefont
  {Embon}}, \bibinfo {author} {\bibfnamefont {Y.}~\bibnamefont {Anahory}},
  \bibinfo {author} {\bibnamefont {\u{Z}. L.~Jeli\'{c}}}, \bibinfo {author}
  {\bibfnamefont {E.~O.}\ \bibnamefont {Lachman}}, \bibinfo {author}
  {\bibfnamefont {Y.}~\bibnamefont {Myasoedov}}, \bibinfo {author}
  {\bibfnamefont {M.~E.}\ \bibnamefont {Huber}}, \bibinfo {author}
  {\bibfnamefont {G.}~\bibnamefont {Mikitik}}, \bibinfo {author} {\bibfnamefont
  {A.~V.}\ \bibnamefont {Silhanek}}, \bibinfo {author} {\bibfnamefont
  {M.~V.~M.}\ \bibnamefont {\u{s}evi\'{c}}}, \bibinfo {author} {\bibnamefont
  {aa. Gurevich}}, \ and\ \bibinfo {author} {\bibfnamefont {E.}~\bibnamefont
  {Zeldov}},\ }\href {\doibase 10.1038/s41467-017-00089-3} {\bibfield
  {journal} {\bibinfo  {journal} {Nature Communications}\ }\textbf {\bibinfo
  {volume} {8}},\ \bibinfo {pages} {85} (\bibinfo {year}
  {2017}{\natexlab{b}})}\BibitemShut {NoStop}%
\bibitem [{\citenamefont {L\"uscher}\ \emph {et~al.}(2007)\citenamefont
  {L\"uscher}, \citenamefont {Milstein},\ and\ \citenamefont
  {Sushkov}}]{Luscher2007}%
  \BibitemOpen
  \bibfield  {author} {\bibinfo {author} {\bibfnamefont {A.}~\bibnamefont
  {L\"uscher}}, \bibinfo {author} {\bibfnamefont {A.~I.}\ \bibnamefont
  {Milstein}}, \ and\ \bibinfo {author} {\bibfnamefont {O.~P.}\ \bibnamefont
  {Sushkov}},\ }\href {\doibase 10.1103/PhysRevLett.98.037001} {\bibfield
  {journal} {\bibinfo  {journal} {Phys. Rev. Lett.}\ }\textbf {\bibinfo
  {volume} {98}},\ \bibinfo {pages} {037001} (\bibinfo {year}
  {2007})}\BibitemShut {NoStop}%
\bibitem [{\citenamefont {Kolkowitz}\ \emph {et~al.}(2015)\citenamefont
  {Kolkowitz}, \citenamefont {Safira}, \citenamefont {High}, \citenamefont
  {Devlin}, \citenamefont {Choi}, \citenamefont {Unterreithmeier},
  \citenamefont {Patterson}, \citenamefont {Zibrov}, \citenamefont
  {Manucharyan}, \citenamefont {Park},\ and\ \citenamefont
  {Lukin}}]{Kolkowitz2015}%
  \BibitemOpen
  \bibfield  {author} {\bibinfo {author} {\bibfnamefont {S.}~\bibnamefont
  {Kolkowitz}}, \bibinfo {author} {\bibfnamefont {A.}~\bibnamefont {Safira}},
  \bibinfo {author} {\bibfnamefont {A.~A.}\ \bibnamefont {High}}, \bibinfo
  {author} {\bibfnamefont {R.~C.}\ \bibnamefont {Devlin}}, \bibinfo {author}
  {\bibfnamefont {S.}~\bibnamefont {Choi}}, \bibinfo {author} {\bibfnamefont
  {Q.~P.}\ \bibnamefont {Unterreithmeier}}, \bibinfo {author} {\bibfnamefont
  {D.}~\bibnamefont {Patterson}}, \bibinfo {author} {\bibfnamefont {A.~S.}\
  \bibnamefont {Zibrov}}, \bibinfo {author} {\bibfnamefont {V.~E.}\
  \bibnamefont {Manucharyan}}, \bibinfo {author} {\bibfnamefont
  {H.}~\bibnamefont {Park}}, \ and\ \bibinfo {author} {\bibfnamefont {M.~D.}\
  \bibnamefont {Lukin}},\ }\href {http://dx.doi.org/10.1126/science.aaa4298}
  {\bibfield  {journal} {\bibinfo  {journal} {Science}\ }\textbf {\bibinfo
  {volume} {347}},\ \bibinfo {pages} {1129} (\bibinfo {year}
  {2015})}\BibitemShut {NoStop}%
\bibitem [{\citenamefont {Kumar}\ \emph {et~al.}(2017)\citenamefont {Kumar},
  \citenamefont {Chandran},\ and\ \citenamefont {Ramachandra~Rao}}]{Kumar2017}%
  \BibitemOpen
  \bibfield  {author} {\bibinfo {author} {\bibfnamefont {D.}~\bibnamefont
  {Kumar}}, \bibinfo {author} {\bibfnamefont {M.}~\bibnamefont {Chandran}}, \
  and\ \bibinfo {author} {\bibfnamefont {M.~S.}\ \bibnamefont
  {Ramachandra~Rao}},\ }\href {https://doi.org/10.1063/1.4982591} {\bibfield
  {journal} {\bibinfo  {journal} {Applied Physics Letters}\ }\textbf {\bibinfo
  {volume} {110}},\ \bibinfo {pages} {191602} (\bibinfo {year}
  {2017})}\BibitemShut {NoStop}%
\bibitem [{\citenamefont {Doherty}\ \emph
  {et~al.}(2014{\natexlab{b}})\citenamefont {Doherty}, \citenamefont
  {Struzhkin}, \citenamefont {Simpson}, \citenamefont {McGuinness},
  \citenamefont {Meng}, \citenamefont {Stacey}, \citenamefont {Karle},
  \citenamefont {Hemley}, \citenamefont {Manson}, \citenamefont {Hollenberg},\
  and\ \citenamefont {Prawer}}]{Doherty2014Pressure}%
  \BibitemOpen
  \bibfield  {author} {\bibinfo {author} {\bibfnamefont {M.~W.}\ \bibnamefont
  {Doherty}}, \bibinfo {author} {\bibfnamefont {V.~V.}\ \bibnamefont
  {Struzhkin}}, \bibinfo {author} {\bibfnamefont {D.~A.}\ \bibnamefont
  {Simpson}}, \bibinfo {author} {\bibfnamefont {L.~P.}\ \bibnamefont
  {McGuinness}}, \bibinfo {author} {\bibfnamefont {Y.}~\bibnamefont {Meng}},
  \bibinfo {author} {\bibfnamefont {A.}~\bibnamefont {Stacey}}, \bibinfo
  {author} {\bibfnamefont {T.~J.}\ \bibnamefont {Karle}}, \bibinfo {author}
  {\bibfnamefont {R.~J.}\ \bibnamefont {Hemley}}, \bibinfo {author}
  {\bibfnamefont {N.~B.}\ \bibnamefont {Manson}}, \bibinfo {author}
  {\bibfnamefont {L.~C.~L.}\ \bibnamefont {Hollenberg}}, \ and\ \bibinfo
  {author} {\bibfnamefont {S.}~\bibnamefont {Prawer}},\ }\href {\doibase
  10.1103/PhysRevLett.112.047601} {\bibfield  {journal} {\bibinfo  {journal}
  {Phys. Rev. Lett.}\ }\textbf {\bibinfo {volume} {112}},\ \bibinfo {pages}
  {047601} (\bibinfo {year} {2014}{\natexlab{b}})}\BibitemShut {NoStop}%
\bibitem [{\citenamefont {Drozdov}\ \emph {et~al.}(2015)\citenamefont
  {Drozdov}, \citenamefont {Eremets}, \citenamefont {Troyan}, \citenamefont
  {Ksenofontov},\ and\ \citenamefont {Shylin}}]{Drozdov2015}%
  \BibitemOpen
  \bibfield  {author} {\bibinfo {author} {\bibfnamefont {A.~P.}\ \bibnamefont
  {Drozdov}}, \bibinfo {author} {\bibfnamefont {M.~I.}\ \bibnamefont
  {Eremets}}, \bibinfo {author} {\bibfnamefont {I.~A.}\ \bibnamefont {Troyan}},
  \bibinfo {author} {\bibfnamefont {V.}~\bibnamefont {Ksenofontov}}, \ and\
  \bibinfo {author} {\bibfnamefont {S.~I.}\ \bibnamefont {Shylin}},\ }\href
  {http://dx.doi.org/10.1038/nature14964} {\bibfield  {journal} {\bibinfo
  {journal} {Nature}\ }\textbf {\bibinfo {volume} {525}},\ \bibinfo {pages} {73
  EP } (\bibinfo {year} {2015})}\BibitemShut {NoStop}%
\bibitem [{\citenamefont {Gor'kov}\ and\ \citenamefont
  {Kresin}(2018)}]{Gorkov2018}%
  \BibitemOpen
  \bibfield  {author} {\bibinfo {author} {\bibfnamefont {L.~P.}\ \bibnamefont
  {Gor'kov}}\ and\ \bibinfo {author} {\bibfnamefont {V.~Z.}\ \bibnamefont
  {Kresin}},\ }\href {\doibase 10.1103/RevModPhys.90.011001} {\bibfield
  {journal} {\bibinfo  {journal} {Rev. Mod. Phys.}\ }\textbf {\bibinfo {volume}
  {90}},\ \bibinfo {pages} {011001} (\bibinfo {year} {2018})}\BibitemShut
  {NoStop}%
\end{thebibliography}

%

\end{document}